\documentclass[aps,prx,reprint,twocolumns,nofootinbib,superscriptaddress]{revtex4-2}

\usepackage{amsmath}
\usepackage{amssymb}
\usepackage{mathrsfs}
\usepackage[dvipsnames]{xcolor}
\usepackage{graphicx}
\usepackage{times}
\usepackage[pdfauthor={Filippo Camilloni}]{hyperref}
\usepackage{float}

\usepackage{booktabs} 

\newcommand{\eg}{e.g., ~}
\newcommand{\ie}{i.e., ~}

\hypersetup{colorlinks=true, 
  linkcolor=blue, 
  citecolor=cyan 
}

\begin{document}

\title{Universality of the Blandford-Znajek emission in stationary and
  axisymmetric spacetimes}
\author{F. Camilloni}
\affiliation{Institut f{\"u}r Theoretische Physik, Max-von-Laue-Strasse
  1, 60438 Frankfurt, Germany}

\author{L. Rezzolla}
\affiliation{Institut f{\"u}r Theoretische Physik, Max-von-Laue-Strasse
  1, 60438 Frankfurt, Germany}
\affiliation{School of Mathematics, Trinity College, Dublin 2, Ireland}
\affiliation{CERN, Theoretical Physics Department, 1211 Geneva 23, 
  Switzerland}

\begin{abstract}
  The Blandford-Znajek (BZ) mechanism is widely recognised as the most
  compelling process to extract rotational energy from an accreting black
  hole and power the emission of relativistic jets. We explore the
  universality of this process for generic black-hole spacetimes within
  the Konoplya-Rezzolla-Zhidenko formalism and find that the lowest-order
  contribution to the BZ power is invariant across different black-hole
  spacetimes. We also show that at the next-leading-order, different
  black-hole spacetimes will lead to different BZ luminosities. As a
  result, while slowly rotating black holes cannot be distinguished via
  measurements of their jet power, rapidly rotating ones have the
  potential of providing information on the strong-field properties of
  the spacetime when independent measurements of the BZ luminosity and of
  the black-hole angular velocity are available.
\end{abstract}

\maketitle

%
\section{Introduction}
\label{sec:intro}

The last decade has marked a series of impressive achievements in
relativistic astrophysics. Some of the most compelling evidences in
support of general relativity (GR) have been provided with the detection
of gravitational waves emitted by coalescing binary systems of black
holes (BHs) and neutron stars~(see \eg~\cite{Abbott2016a, LIGOGW170817}), 
as well as with the direct observations of accreting supermassive BHs in
galactic centres~\cite{EventHorizonTelescope2019_L1, 
  EventHorizonTelescope2022_L1}. These observations critically
contributed to reinforce the acceptance of the existence of BHs and their
role in some of the most powerful astrophysical phenomena. Such a picture
is further enriched by the possibility envisioned by
Penrose~\cite{Penrose1971} that spinning BHs constitute tremendous
reservoirs of rotational energy available to be extracted. Since this
original insight, energy-extraction processes have been studied under a
variety of physical conditions~\cite{Press1972, Bardeen72, Denardo1973, 
  Wagh1985, Williams1995, Koide2003, Koide2008, Banados2009, 
  Schnittman2015, Brito2015, East2017, Parfrey2019, Comisso2021, 
  Meringolo2025a, Camilloni2025}. The Blandford-Znajek (BZ)
mechanism~\cite{Blandford1977}, in particular, stands for his importance
as the driving engine behind the formation of relativistic jets in
accreting BHs. Indeed, the BZ mechanism played a crucial role in
providing a unified explanation to how relativistic jets are launched and
sustained by spinning BHs across all mass-scales: from stellar mass BHs
produced by binary neutron star mergers, source of gravitational waves
followed by jets and short gamma-ray bursts \cite{Abramowicz2013, 
  Ruiz2021rev}, to supermassive BHs in active galactic
nuclei~\cite{Narayan2012b}, that launch extremely energetic relativistic
jets, propagating steadily beyond galactic lengthscales.

The great success of this model in explaining jet emission is related to
the minimal set of conditions required for it to operate, that can be
naturally achieved in a number of astrophysical scenarios. Indeed, the BZ
mechanism can be active whenever a spinning BH is immersed in a strongly
magnetised and plasma-rich environment. It is natural to expect these
conditions to be met as a result of neutron stars binary
mergers~\cite{Eichler89, Narayan92, Rezzolla:2011, Palenzuela2015, 
  Ruiz2016, Ciolfi2020c, East2021, Hayashi2025}, whereas the Event
Horizon Telescope (EHT) collaboration provided crucial evidences for
their validity around accreting supermassive
BHs~\cite{EventHorizonTelescope2021_L7, EventHorizonTelescope2021_L8, 
  EventHorizonTelescope2024_L7, EventHorizonTelescope2024_L8}. When the
BZ mechanism is active, part of the BH rotational energy is transferred
to an ordered large-scale magnetosphere and it is efficiently converted
into Poynting fluxes and strong poloidal currents that accelerate
particles along the magnetic field lines at relativistic speeds.
Although the theoretical predictions of the power of relativistic jets
are affected by uncertainties related to plasma, kinetic and
magnetohydrodynamical processes, as well as nonlinearities introduced
by the environment, it is customary to regard the BZ power as a proxy
for the jet luminosity. The two powers have been shown to be
consistent not only for persistent jets in active galactic
nuclei~\cite{Ghisellini2014}, but also for transient ballistic jets
launched by X-rays binaries~\cite{Narayan2012b}. Hereafter, we will
adopt this strict correlation between these two quantities, so that 
the measurement of the BZ luminosity constitutes an opportunity to
obtain strong-gravity information about the environment in which
relativistic jets originate.

Astrophysical BHs in GR are described by the Kerr spacetime, a
two-parameters family of solutions specified by the BH mass and
angular momentum. In their work~\cite{Blandford1977}, BZ derived a
characteristic quadratic scaling for the jet luminosity as a function
of the angular velocity of a Kerr BH. More recently, both
numerical~\cite{Tchekhovskoy2010, Meringolo2025a} and
analytic~\cite{Tanabe2008, Camilloni2022} studies converged to more
refined expressions for the BZ luminosity, in which the original
quadratic scaling receives additional corrections that become relevant
for Kerr BHs in the high-spin regime. On the other hand, a number of
alternative theories and possible completions of GR have lead to an
impressive number of new BH scenarios that deviate from the Kerr
metric. A non-exhaustive list of examples range from BHs emerging in
the low-energy limit of string-theory~\cite{Sen1992}, hairy
BHs~\cite{Herdeiro2014}, regular BHs~\cite{Bardeen68, Hayward2005}, as
well as BHs in modified gravity~\cite{Yunes2009, Alexander:2009tp}.

Given these considerations, it becomes relevant to explore whether the BZ
luminosity can bear imprints of the background spacetime and if such
information can be used to test the Kerr hypothesis~\cite{Bambi2012a}.
Our goal in this work is to answer the following question: ``Under
which conditions, and to what extent, is it \emph{in principle}
possible to distinguish BHs in different theories of gravity by
measuring their jet luminosity?''
In addition to what done so far in terms of horizon-scale imaging
(see, \eg \cite{Younsi2016, Mizuno2018, Uniyal2025}), previous studies
have already partially considered this problem for specific scenarios
that include extensions of GR~\cite{Banerjee2020, Dong2021,
  Chanson2022, Chatterjee2023, Peng2023, Yang2025a}, alternative
theories of gravity~\cite{Pei2016, Armengol2016, Camilloni2023} and
even wormholes~\cite{Urtubey2025}.
Here, we follow a different logic and rather than exploring yet
another representative spacetime out of the large variety of possible
scenarios, we develop a generic and theory-agnostic framework to study
the BZ mechanism. Being our approach essentially analytic, we can in
principle cover the entire parameter space and bypass the daunting
task of running several numerical simulations for each choice of the
initial parameters.

More specifically, we consider the BH magnetospheric problem and the
emission of relativistic jets via the BZ mechanism by solving
perturbatively the system of equations of
\emph{force-free-electrodynamics} (FFE) in a parameterised background
given by the \emph{Konoplya-Rezzolla-Zhidenko} (KRZ)
metric~\cite{Konoplya2016a}, which extends to stationary and axisymmetric
spacetimes the \emph{Rezzolla-Zhidenko} (RZ) parameterisation of static
spacetimes~\cite{Rezzolla2014}. This choice is motivated by the superior
convergence properties of the KRZ approach with respect to other
parametric frameworks~\cite{Johannsen2011, Johannsen2013, Cardoso2014},
and enables one to build good approximations for many of the known BH
solutions by fixing a small set of parameters.

In this way, we obtain an analytic and theory-agnostic expression for the
power extracted in the BZ mechanism that is of 6th-order in the BH
angular velocity $\Omega_h$ and that paves the way to further analytic
and numerical explorations\footnote{We should note that a previous
attempt has been made to study the BZ luminosity using the KRZ
metric~\cite{Konoplya2021}. While an important first step, this work was
limited to the second order in $\Omega_h$ and adopted a sub-optimal set
of horizon-penetrating coordinates.}. Such expression reveals a
\emph{quasi-universal} nature of the leading-order behaviour, namely,
that the BZ luminosity scales as a quadratic function of the BH angular
velocity, \ie $P_{_{\rm BZ}} \sim \Omega_h^2$. As we will show, this
result is non-trivial and points out to the intrinsic difficulty of
distinguishing different theories of gravity by measuring the luminosity
of jets launched from BHs that are slowly-rotating. While the same
degeneracy has been previously noted in specific alternative theories of
gravity~\cite{Dong2021, Camilloni2023}, the use of the KRZ framework is
what enabled us to assess the quasi-universality of this
result. Fortunately, however, the mathematical structure of the agnostic
BZ luminosity is such that it can be cast into a generic form
\hbox{$(P_{_{\rm BZ}})_{_{\rm KRZ}} \sim \Omega_h^2\, f_{_{\rm
      KRZ}}(\Omega_h)$}~\cite{Tchekhovskoy2011, Camilloni2022,
  Meringolo2025a}, where $f_{_{\rm KRZ}}(\Omega_h)$ is computed up to
$\mathcal{O}(\Omega^4_h)$ and it depends on the details of the background
spacetime.
As a result, for BHs that are rapidly rotating, the BZ jet potentially
enabling tests of the Kerr paradigm in future high-precision
horizon-scale observations of relativistic jets. To this scope, we
provide analytic approximations for the high-spin corrections
$f_{_{\rm KRZ}}(\Omega_h)$ so as to produce the first catalogues of BZ jet
luminosity curves for different BHs within the KRZ framework.

The article is structured as follows. Section~\ref{sec:KRZ} is devoted
to introduce the KRZ framework, study the physical constraints on the
parameter space and classify a spacetime classification based on the
BH angular velocity. Our main physical results are presented in
Sec.~\ref{sec:ffe}, where we compute the theory-agnostic expression
for the BZ luminosity, present explicit cases and construct a
catalogue of BZ curves. A self-consistent formulation of the BH
magnetospheric problem and the BZ mechanism in KRZ spacetimes is
provided in App.~\ref{app:BHMAG}, \ref{app:BZ} and ~\ref{sec:ffepert},
together with many mathematical details and ancillary results.

\section{Theory-agnostic black hole metric}
\label{sec:KRZ}

Because the fundamental goal of our study is to assess the universality
of the BZ mechanism across different BH spacetimes, an essential first
step lies in employing a description that is both generic and effective
in reproducing known BH solutions, without the constraint of wishing to
reproduce specific properties of the solutions in GR (see, \eg
Ref.~\cite{Achour2025} for a recent approach where a closer connection
with the Kerr solution is instead imposed). While several avenues are
possible, our choice here has fallen on the Konoplya-Rezzolla-Zhidenko
(KRZ) metric parameterisation~\cite{Konoplya2016a}. The KRZ framework was
developed as a generalisation to axially symmetric BH spacetimes of the
Rezzolla-Zhidenko (RZ) parameterisation, originally developed for static
and spherically symmetric spacetimes and from which it derives much of
its mathematical properties~\cite{Rezzolla2014}. The KRZ framework
permits a theory-agnostic characterisation of stationary BH spacetimes in
arbitrary metric theories, overcoming many of the difficulties of
previous approaches~\cite{Johannsen2011, Johannsen2013, Cardoso2014}, and
successfully converging to various BH metrics with few deformation
parameters~\cite{Younsi2016}. Besides providing reference BH metrics to
compare with imaging of supermassive BHs~\cite{Kocherlakota2020,
  EHT_SgrA_PaperVI}, the RZ parameterisation has also been employed to
derive generic equilibrium solution of non-self-gravitating fluid
tori~\cite{Cassing2023}.

While the original version of the KRZ metric is properly defined only
outside the event horizon, an important feature of BH spacetimes is that
the event horizon must constitute a removable (\ie coordinate)
singularity. Moreover, in order to avoid complicated boundary conditions
at the horizon, numerical simulations usually exploit horizon-penetrating
(HP) coordinates that allow analytical continuations of the computational
domain in the BH interior, where excision can be safely performed due to
its causally-disconnected nature. These considerations have recently
pushed the research towards novel formulations of the KRZ metric in HP
coordinates, that can smoothly cross the horizon and extend to the BH
inner region~\cite{Konoplya2021, Ma2024}.

In this context, the choice that best adapts to our aims, and that will
facilitate future numerical simulations to make contact with our results,
hereafter we will employ the HP form of the KRZ metric proposed in
Ref.~\cite{Ma2024}, whose line-element reads
\begin{align}
 \label{eq:HP_KRZ_metric}
 &ds^2 =-\left(1 - \frac{R_{_{\rm M}}}{r\Sigma}\right) d t^2 + 2
 \frac{R_{_{\rm M}}R_{_{\rm B}}} {r\Sigma} d t d r \nonumber \\
 &- 2\frac{M R_{_{\rm M}}}{r\Sigma} a_* \sin^2{\theta} d
 t d \phi - 2M \left(1 + \frac{R_{_{\rm M}}}{r \Sigma}\right) R_{_{\rm B}} a_*
 \sin^2{\theta} d r d \phi
 \nonumber\\
 &+ \left(1 + \frac{R_{_{\rm M}}}{r\Sigma}\right) R_{_{\rm B}}^2
 d r^2+ r^2 \Sigma d \theta^2 + r^2 K^2
 \sin^2{\theta} d \phi^2 \, , 
\end{align}
and where $M$ and $a_*$ are respectively the BH mass and its
dimensionless spin parameter. The metric determinant is given by $g :=
-R_{_{\rm B}}^2 \Sigma^2 r^4 \sin^2{\theta}$, and the explicit expression
for the metric functions are
\begin{equation}
  \begin{split}
    \Sigma(r, \theta) &:= 1 +a_*^2 \frac{ M^2\cos^2{\theta}}{r^2}\, , 
    \\
    K^2(r, \theta) &:= \Sigma(r, \theta) +a_*^2 \left(1 +
 \frac{R_{_{\rm M}}(r)}{r\Sigma(r, \theta)}\right) \frac{
   M^2 \sin^2{\theta}}{r^2}\, .
 \end{split}
\end{equation}
The position of the event horizon, that is conventionally labelled as
$r=r_0$, is specified by the largest root of the equation 
\begin{equation}
  \label{eq:N2.eq.0}
  N^2(r, \theta) := 1 - \frac{R_{_{\rm M}}(r)}{r} +
 a_*^2\frac{M^2}{r^2} = 0\, , 
\end{equation}
whereas the ergosphere radius is determined by the solution of $\Sigma(r,
\theta)=R_{_{\rm M}}(r)/r$. The functional dependence of the two
functions $R_{_{\rm M}}$ and $R_{_{\rm B}}$ is what ultimately
characterise the KRZ framework and introduces the dependence on
additional spacetime parameters.

The KRZ formalism is based on Pad\'e approximants in the form of
continued fractions that ensure superior convergence
properties~\cite{Rezzolla2014, Konoplya2016a}. This allows the metric
\eqref{eq:HP_KRZ_metric} to capture deviations from Kerr spacetimes with
same mass and angular momentum by using a limited number of
parameters~\cite{Kocherlakota2020, Younsi2016}.
More specifically, the two functions $ R_{_{\rm B}}$ and $ R_{_{\rm M}}$
contain the information about the deformation parameters via the
following expressions~\cite{Konoplya2018} expressed in terms of the
compactified radial coordinate $\tilde{x}:= 1-r_0/r$
\begin{align}
  \label{eq:RB_expan}
  R_{_{\rm B}} &= 1 + b_0 (1 - \tilde{x}) + \dfrac{b_1 (1 - \tilde{x})^2}{1 +
   \dfrac{b_2 \tilde{x}}{1 + \dfrac{b_3 \tilde{x}}{1 + \cdots}}}\, , 
 \\\nonumber
 \label{eq:RM_expan}
 \frac{R_{_{\rm M}}}{r} &= 1 - \tilde{x} + a_*^2\frac{M^2}{r_0^2} (1 -
 \tilde{x})^3 + \epsilon \tilde{x} (1 - \tilde{x}) \\ &\quad- (a_0 -
 \epsilon) \tilde{x} (1 - \tilde{x})^2- \dfrac{a_1 \tilde{x} (1 -
   \tilde{x})^3}{1 + \dfrac{a_2 \tilde{x}}{1 + \dfrac{a_3 \tilde{x}}{1
       + \cdots}}}\, , 
\end{align}
such that $\tilde{x}=0$ and $\tilde{x}=1$ correspond to the event horizon
and to spatial infinity, respectively. The parameters $\epsilon$, $a_0$
and $b_0$ determine the asymptotic properties, whereas $a_{1, 2, \dots}$
and $b_{1, 2, \dots}$ determine the near-horizon properties.

An important parameter in the RZ parameterisation was introduced to
measure the possible deviation of the position of the event horizon from
the general-relativistic value of the Schwarzschild radius, \ie
\begin{equation}
\epsilon := \frac{2 M}{r_0} - 1\, ,
\end{equation}
While such a definition simplifies the comparison with parameterised
post-Newtonian approach at large distances~\cite{Rezzolla2014}, it needs
an extension when considering the location of the event horizon in an
axisymmetric spacetime. In such a case, in fact, the position of the
event horizon will depend also on the spin of the black hole and hence
the parameter $\epsilon$ needs to be replaced by its extension 
\begin{equation}
  \varrho :=  \frac{\epsilon (1+\sqrt{1-a^2_*})^2 - a^2_*}
  {(1+\sqrt{1-a^2_*})^2 + a^2_*}\, , 
\end{equation}
such that 
\begin{equation}
  \epsilon=\varrho + \frac{a_*^2(1+\varrho)}{(1+\sqrt{1-a^2_*})^2}\, . 
\end{equation}
In this way, the location of the event horizon is given by
\begin{equation}
  \label{eq:r0}
  r_0 := M \left( \frac{1+\sqrt{1-a^2_*}}{1+\varrho} \right) \, ,
\end{equation}
where demanding $r_0$ to be real implies that $0 \leq a_*^2 \leq 1$. In
what follows we assume that $a_* \geq 0$.

Clearly, for any value of $a_*$ the location of the event horizon
coincides with that of the Kerr spacetime when $\varrho=0$, and in
spherical symmetry, \ie for $a_*=0$, $\varrho$ reduces to $\epsilon$, as
expected. In what follows, we will refer to as ``KRZ parameters'' the set
of coefficients $\varrho$, $a_{0, 1, \dots}$ and $b_{0, 1, \dots}$ and
distinguish them from $a_*$ and $M$ that are also parameters of the Kerr
solution. The asymptotic parameters can be interpreted as combinations of
parameterised post-Newtonian quantities, and current observational
constraints allows to set $a_0\approx 0$ and $b_0\approx
0$~\cite{Konoplya2020, Will:2006LRR} as a first reasonable
approximation. To keep the discussion at a level that can be handled
analytically, hereafter we will restrict our attention to the section of
the possible KRZ spacetimes spanned by the space of parameters $(\varrho,
a_1, b_1)$, together with the BH mass and $(M,a_*)$. We note that the
restriction in the parameter space does not affect the generality of the
methods employed, nor the impact of conclusions drawn, since this section
of the space of the solutions already provides a very good approximation
to many of the known BH metrics~\cite{Konoplya2020}.

Having established the section of possible solutions, we next need to
define the boundaries of the sub-space, that is, we need to define the
numerical values that the parameters $\varrho, a_1$, and $b_1$ are
allowed to take when describing physically reasonable solutions. To this
scope, we extend the strategy employed in Ref.~\cite{Kocherlakota2022}
for the static backgrounds to finite values of the BH spin. More
specifically, we demand that the largest root of the
Eq.~\eqref{eq:N2.eq.0} corresponds to the location of the event horizon
$r_0$ as defined by Eq.~\eqref{eq:r0}. Given that the metric function $N$
depends on $R_{_{\rm M}}$, the condition on the BH horizon allows us to
impose constraints over the KRZ parameters $\varrho$ and $a_1$.

\begin{figure}[hb]
  \hspace{-5mm}\includegraphics[width=0.44\textwidth]{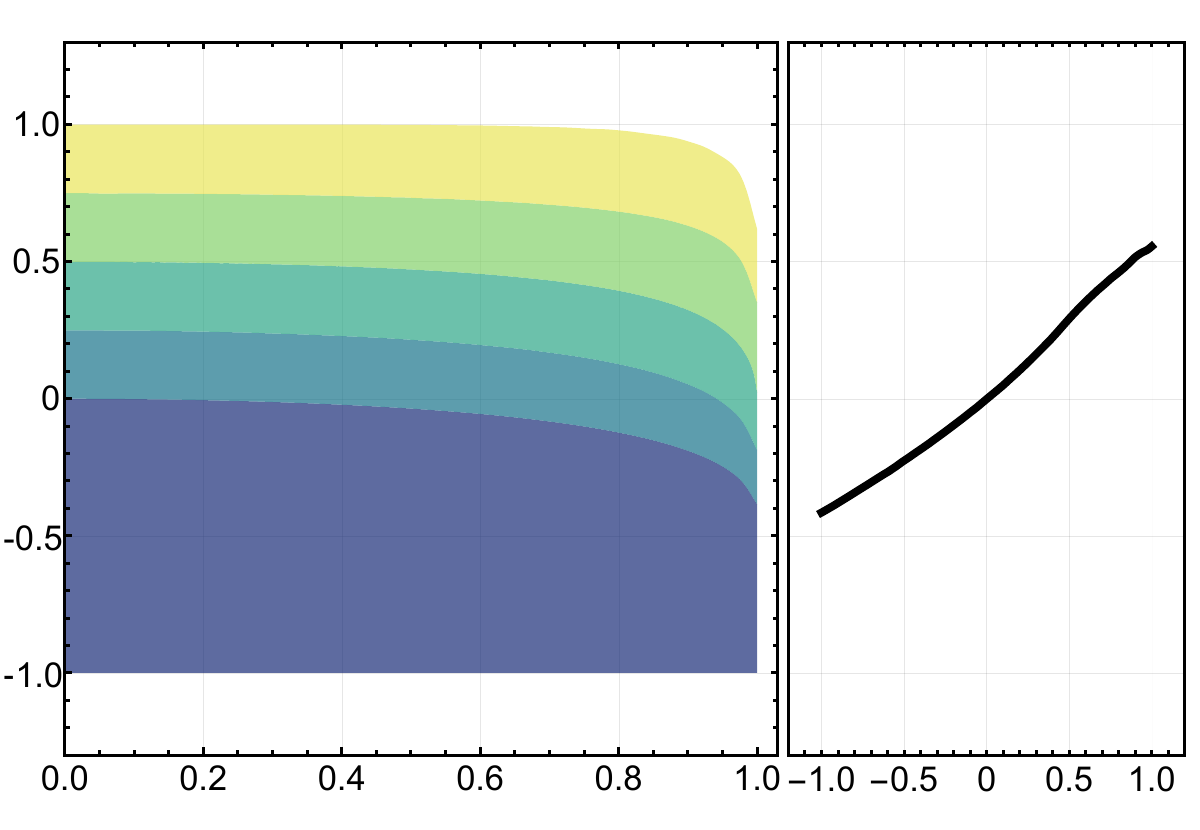}
  \begin{picture}(0, 0)
     \put(-230, 80){\large $\varrho$}
     \put(-120, 122){\scriptsize $a_1=1$}
     \put(-140, 108){\scriptsize $a_1=1/2$}
     \put(-155, 96){\scriptsize $a_1=0$}
     \put(-180, 84){\scriptsize $a_1=-1/2$}
     \put(-200, 72){\scriptsize $a_1=-1$}
     \put(-154, 0){\large $a_*$}
     \put(-33, 83){\rotatebox{0}{\large$\varrho^{_{\rm extr}}$}}
     \put(-45, -2){\large$a_1$}
     \put(-75, 139){\footnotesize $a_*=1$}
  \end{picture}
  \\
   \vspace{5mm}
   \hspace{-5mm}\includegraphics[width=0.44\textwidth]{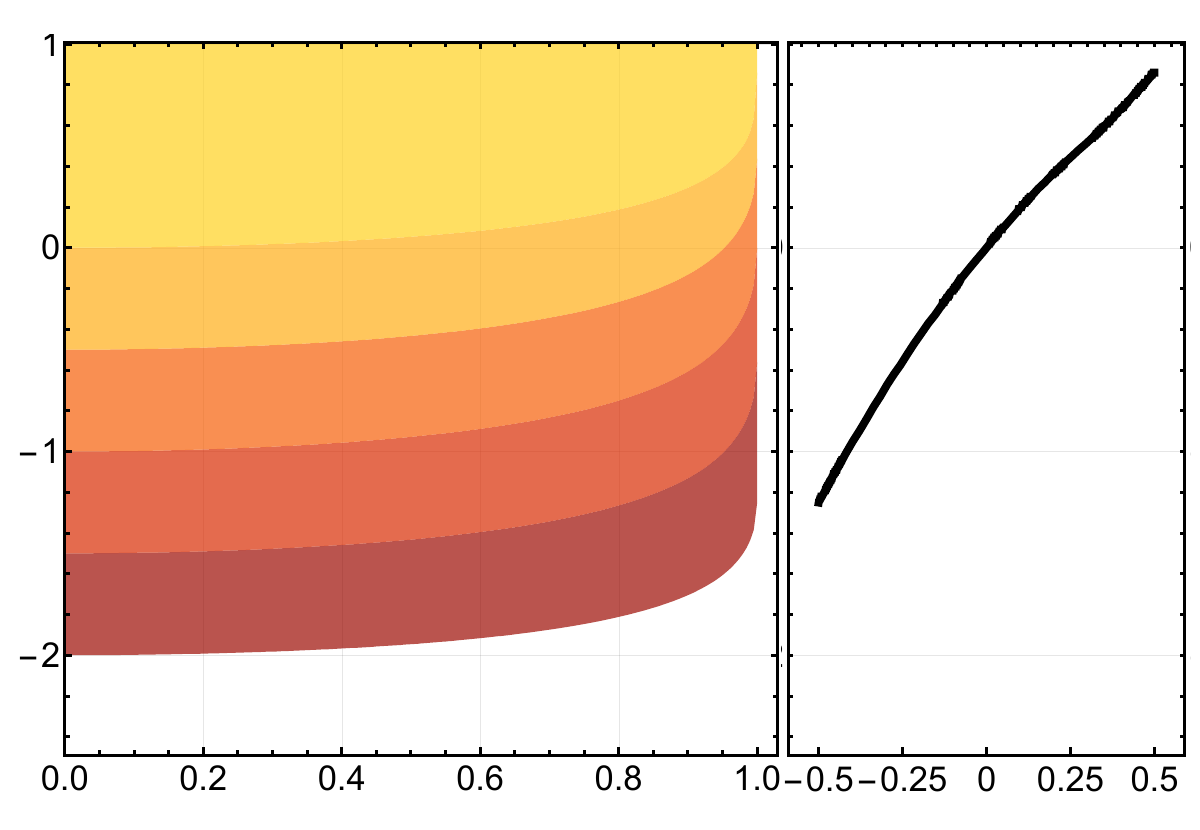}
     \begin{picture}(0, 0)
     \put(-230, 80){\large $a_1$}
     \put(-122, 125){\scriptsize $\varrho=1/2$}
     \put(-140, 104){\scriptsize $\varrho=1/4$}
     \put(-155, 82){\scriptsize $\varrho=0$}
     \put(-180, 61){\scriptsize $\varrho=-1/4$}
     \put(-200, 43){\scriptsize $\varrho=-1/2$}
     \put(-154, -2){\large $a_*$}
     \put(-50, 92){\rotatebox{0}{\large$a_1^{_{\rm extr}}$}}
     \put(-45, -2){\large$\varrho$}
     \put(-75, 139){\footnotesize $a_*=1$}
  \end{picture}
     \caption{\textit{Top panel:} Admissibility region for the KRZ
       parameter $\varrho$ for different values of the BH spin $a_*$ and
       of the $a_1$ parameter. As summarised in
       Eq.~\eqref{eq:KRZconstraints}, for any fixed value of $a_*$,
       $\varrho$ is allowed to range between a maximum value
       $\varrho^{_{\rm max}}$ and a minimum value $a_1^{_{\rm min}}$. The
       plots on the right represent the values of $\varrho^{_{\rm extr}}$
       attained at $a_*=1$. \textit{Bottom panel:} the same as in the top
       one but for the KRZ parameter $a_1$ for different values of $a_*$
       and $\varrho$. The plots on the right represent the values of
       $a_1^{_{\rm extr}}$ attained at $a_*=1$.}
\label{fig:param}
\end{figure}
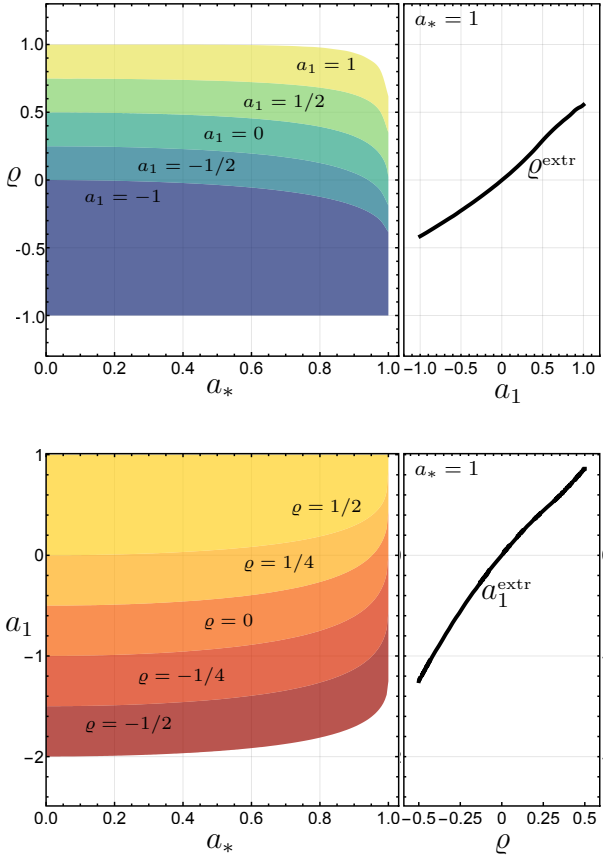

In Fig.~\ref{fig:param} we show the regions where the condition above is
satisfied for the parameters $\varrho$ and $a_1$ by varying the spin
parameter $a_*$. In both cases the values for non-spinning BHs, $a_*=0$,
correspond to the constraints derived in the literature for the RZ metric
(see Table 2 in Ref.~\cite{Kocherlakota2022}). Figure~\ref{fig:param}
shows the existence of an upper, $\varrho^{_{\rm max}}(a_1, a_*)$, and a
lower, $a_1^{_{\rm min}}(\varrho, a_*)$, bound of the physical region.
It is crucial to notice that, because of these bounds, not every BH in
the HP-KRZ class can saturate the extremality condition $a_*=1$. This is
clearly illustrated in Fig.~\ref{fig:param}, where in the left panel it
is easy to observe that lines of constant $\varrho$ and $a_1$ can
intercept the physical bound even for $a_*^{_{\rm max}}<1$. In the right
panel, instead, we show the maximum and minimum values for $\varrho$ and
$a_1$ that a KRZ BH at extremality can possess, \ie $\varrho^{_{\rm
    extr}}:=\varrho^{_{\rm max}}(a_1;a_*=1)$ and $a_1^{_{\rm
    extr}}:=a_1^{_{\rm min}}(\varrho;a_*=1)$. The physical constraints
in the sub-region $(\varrho, a_1)$ of the KRZ parameter space can thus be
summarised as follows
\begin{align}
  \label{eq:KRZconstraints}
  \varrho^{_{\rm max}}(a_1, a_*) &\geq \varrho >-1\, ,\\
  a_1 &\leq a_1^{_{\rm min}}(\varrho, a_*)\, ,\\
  a^{_{\rm max}}_* &\leq 1\, , 
\end{align}
where the curves marking $\varrho^{_{\rm max}}(a_1, a_*=1)$ and $a_1^{_{\rm
    min}}(\varrho, a_*=1)$ are shown in the right panels of
Fig.~\ref{fig:param}.

\subsection{A classification of KRZ spacetimes}
\label{sec:subsuperKerr}

As for every spacetime possessing an intrinsic angular momentum, the KRZ
metric exhibits a dragging of inertial frames, encoded in the angular
velocity
\begin{equation}
  \omega(r,\theta) :=-\frac{g^{r\phi}}{g^{rt}}=\frac{a_*M}{r R_{_{\rm M}}}\,.
\end{equation}
When evaluated at the KRZ horizon, $r=r_0$, this angular velocity
provides a definition for the BH (or horizon) angular velocity
\begin{align}
  \label{eq:omegaH}
  \Omega_h &:= \frac{1}{M} \left( \frac{a_*}{a_*^2 + r_0^2/M^2}\right)
  \nonumber \\
  & = \frac{1}{M} \left[\frac{a_*}{a_*^2
    + \left({1 + \sqrt{1-a_*^2}}/({1 + \varrho})\right)^2}\right]\, ,
\end{align}
where the first relation shows the same functional form valid in the Kerr
metric, whereas in the second one we make explicit use of
Eq.~\eqref{eq:r0}. Expression~\eqref{eq:omegaH} highlights that the value
of the BH angular velocity is controlled by how rapidly the BH spins, \ie
by the parameter $a_*$, and by the size of the horizon radius $r_0$,
which, in the KRZ parameterisation, depends on the parameter $\varrho$.
As a result, the horizon angular frequency is degenerate with respect to
the KRZ parameters, so that two different metrics characterised by
distinct values of $a_{0, 1, \dots}$ or $b_{0, 1, \dots}$ will have the
same horizon angular velocity if the spin and the horizon radius are the
same. As we discuss below, this has profound implications for the BZ
luminosity.\\

\begin{figure}[t]
  \centering
 \includegraphics[width=0.4\textwidth]{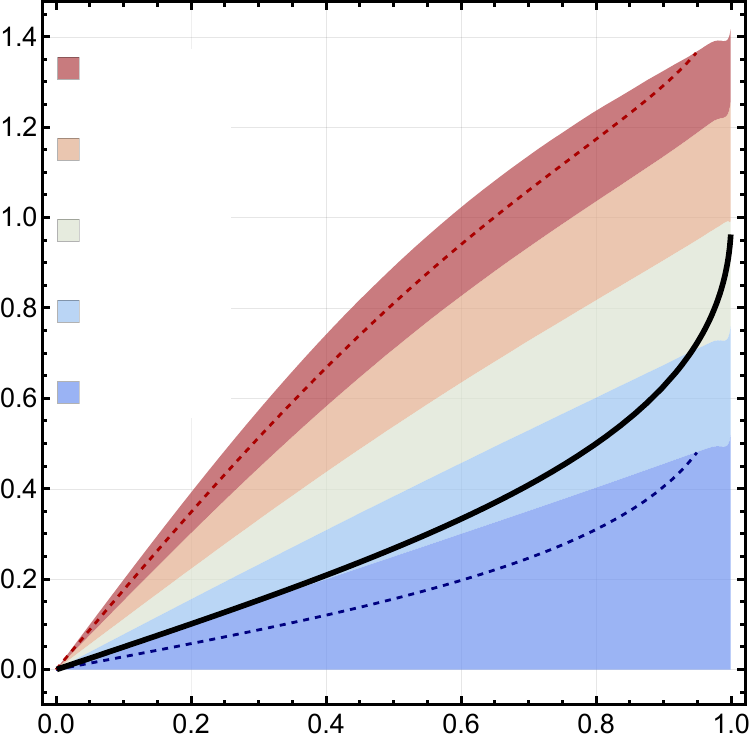}
   \begin{picture}(0, 0)
     \put(-179, 181){\small $a_1=1$}
     \put(-179, 159){\small $a_1=1/2$}
     \put(-179, 137){\small $a_1=0$}
     \put(-179, 115){\small $a_1=-1/2$}
     \put(-179, 92){\small $a_1=-1$}
     \put(-45, 157){\small \rotatebox{35}{\color{BrickRed}{\small $\varrho=0.882$}}}
     \put(-50, 48){\small \rotatebox{28}{\color{Blue}{\small $\varrho=-0.247$}}}
     \put(-218, 95){\large\rotatebox{90}{$2M\Omega_h$}}
     \put(-100, -5){\large$a_*$}
     \put(-105, 39){\rotatebox{20}{\large ${\rm sub-Kerr}$}}
     \put(-115, 51){\rotatebox{25}{\large ${\rm super-Kerr}$}}
   \end{picture}
  \caption{Range of variation of the BH angular velocity $\Omega_h$ as
    a function of the BH spin $a_*$ in the space allowed by changes in
    the KRZ parameter $\varrho$. Shown with a black solid line is the
    well-known nonlinear relation for a Kerr BH, \ie for
    $\varrho=0$. The line distinguishes the possible solutions in a
    super-Kerr regime (above the solid line) and in a sub-Kerr one
    (below the solid line). Shown with different colours are regions
    with constant values of the KRZ parameter $a_1$ that enters in
    setting the range for $\varrho$ (see top panel of
    Fig.~\ref{fig:param}). Finally, reported with a red and blue
    dotted line are representative sub/super-Kerr relations $\Omega_h
    = \Omega_h(a_*)$ for selected values of the parameter $\varrho$
    such that $\varrho=\varrho^{_{\rm max}}$ for $a_*^{_{\rm max}}=0.95$.}
\label{fig:omega}
\end{figure}
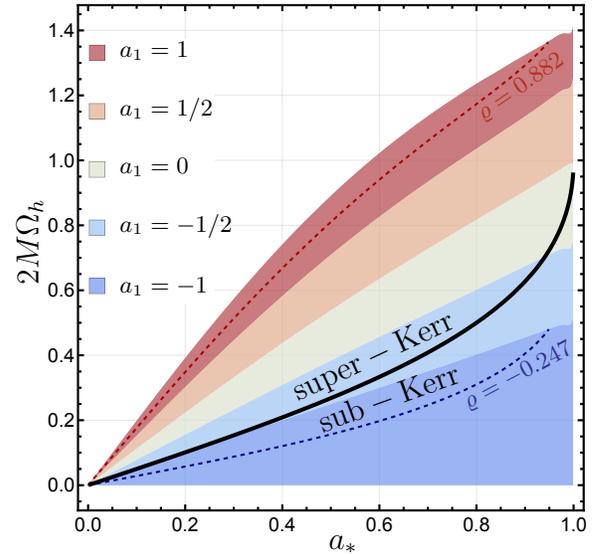

Using the definitions~\eqref{eq:r0} and~\eqref{eq:omegaH} we can compare
the values of $r_0$ and $\Omega_h$ with the corresponding quantities in a
Kerr spacetime, \ie $\bar{r}_0$ and $\bar{\Omega}_h$. In making such
comparisons, it is important to recall that the KRZ parameters are
subject to physical constraints. In particular $\varrho^{_{\rm max}} \geq
\varrho>-1$, where $\varrho^{_{\rm max}}$ depends on the BH spin $a_*$, as
well as on the KRZ parameters $a_{0, 1, \dots}$, so that this also
impacts on the possible values that $\Omega_h$ can take.

Figure~\ref{fig:omega} presents such a comparison when varying the
parameters $\varrho$ and $a_1$ and this allows us distinguish the
following classes of spacetimes
\begin{itemize}
\item[-] \emph{sub-Kerr spacetimes} when $0>\varrho>-1$. In this case,
  for every BH spin $a_*$, the angular velocity is always smaller than
  that of a Kerr BH, \ie $\Omega_h/\bar{\Omega}_h < 1$, since horizon
  radius is always larger than the corresponding radius in Kerr $r_0 >
  \bar{r}_0$.
\item[-] \emph{quasi-Kerr spacetimes} with $\varrho=0$ but any of $a_{0,
  1}\dots$ $b_{0, 1}\dots$ non-vanishing. In this case, both the horizon
  radius and the angular velocity are the same of a Kerr BH with the same
  spin, \ie $\Omega_h/\bar{\Omega}_h = 1$. Note that the spacetimes can
  still significantly differ from Kerr, as indicated by the Kretschmann
  scalar (see Fig.~\ref{fig:delta} and discussion below).
\item[-] \emph{super-Kerr spacetimes} for $\varrho^{_{\rm
    max}} \geq \varrho>0$. In this case, the KRZ angular velocity
  exceeds the corresponding value for a Kerr BH with same $a_*$, \ie
  $\Omega_h /\bar{\Omega}_h > 1$, since $r_0 < \bar{r}_0$. When
  $\varrho>0$ and $a_1>0$ the angular velocity can also exceed the
  extreme value of the Kerr spacetime, $\Omega_h > 1/2M$ and for a
  fixed $\tilde{\varrho}$ there exists a corresponding maximum value
  of the BH spin $a_* \leq \tilde{a}^{_{\rm max}}_*\leq1$ such that
  $\varrho^{_{\rm max}}(\tilde{a}^{_{\rm max}}_*) =:
  \tilde{\varrho}$. For these spacetimes, certain combinations of the
  KRZ parameters reduce the effective maximum value of the BH spin.
\end{itemize}
The rationale for this classification in terms of the ratio
$\Omega_h/\bar{\Omega}_h$ is rooted in the functional dependence of
the BZ power from powers of $\Omega_h$~\cite{Tchekhovskoy2011,
  Camilloni2022}, so that we will later on use this classification to
deduce if the BZ mechanism in a given HP-KRZ spacetime is more or less
efficient that in a Kerr spacetime. Furthermore, we expect that this
classification of the KRZ parameterisation will also help future
studies of the astrophysical phenomenology associated with these
spacetimes.

\subsection{Quantifying the spacetime-deviation from Kerr}
\label{sec:delta}

The relative classification of the KRZ spacetimes discussed above,
naturally leads to the possibility of quantifying the deviation
introduced by a given value of the coefficients in the parameterisation
from the Kerr solution. To this scope, it is useful to compute the
Kretschmann scalar \hbox{$\mathcal{K}_{_{\rm KRZ}} = R^{\mu\nu\rho\sigma}
  R_{\mu\nu\rho\sigma}$} of the HP-KRZ metric as a gauge-invariant
quantity that is dependent by all KRZ parameters. Since this scalar
depends on the specific position in the spacetime and on the black-hole
spin, we compute $\mathcal{K}_{_{\rm KRZ}}$ at the equatorial horizon
(\ie at $r=r_0$ and $\theta=\pi/2$) of a BH spinning at its extremal
value $a_*=1$, and denote this representative quantity as $\kappa_{_{\rm
    KRZ}}$. Furthermore, we measure the gauge-invariant relative
deviation from Kerr as
\begin{equation}
\Delta_{_\%} := 1-\frac{\kappa_{_{\rm KRZ}}}{\kappa} \,.
\end{equation}
Obviously, the definition we adopt for $\Delta_\%$ is not unique, and
other definitions can lead to different values of the spacetime
fractional deviation. Nonetheless, a choice based on the Kretschmann
scalar $\kappa_{_{\rm KRZ}}$ appears as the most natural and as a
representative for the true spacetime curvature (see also
Ref.~\cite{Garibay2026} for a discussion of different curvature
invariants in compact stars). 

\begin{figure}[ht]
  \centering
  \includegraphics[width=0.47\textwidth]{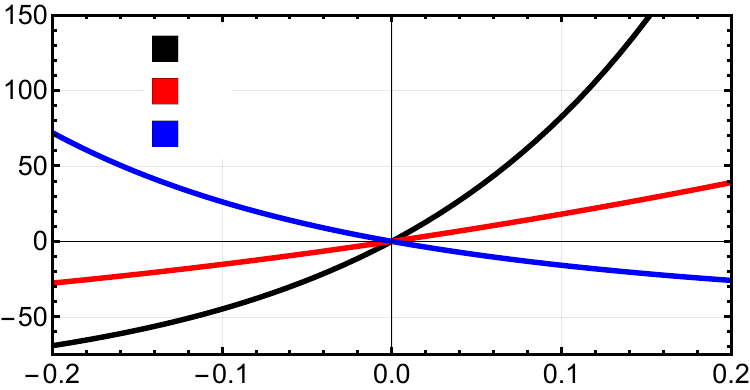}
   \begin{picture}(0, 0)
     \put(-135, -8){\large $\varrho, ~a_1, ~b_1$}
     \put(-145, 125){\rotatebox{0}{\large $\Delta_{\%}(a_*=1)$}}
     \put(-182, 106){\small $\varrho$}
     \put(-182, 93){\small $a_1$}
     \put(-182, 79){\small $b_1$}
   \end{picture}
 \caption{Variation of the relative spacetime deviation $\Delta_\%$
   computed from the Kretschmann scalar at $a_*=1$ as a function of three
   KRZ parameters. Black, red and blue lines refer to variations
   introduced by changes in $\varrho$, $a_1$, and $b_1$, respectively.}
\label{fig:delta}
\end{figure}
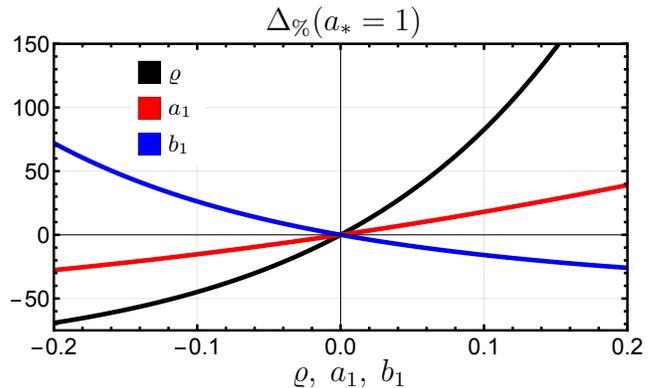

Figure~\ref{fig:delta} reports the fractional deviation for specific
variations of only one of the three KRZ parameters in the range
$(\varrho, a_1, b_1) \in [-0.2, 0.2]$, while keeping the other two equal
to zero. In this way, it is possible to appreciate that even small values
of the KRZ parameters are capable of producing significant spacetime
deviations. In particular, deviations of the order $|\Delta_{_\%}|
\approx \pm 10\%$ are attained already for relatively small reference
values $\varrho \approx \pm 0.02$, $a_1 \approx \pm 0.06$ and $b_1
\approx \mp 0.05$.
\vspace{5mm}
%

\section{Force-Free Electrodynamics in KRZ spacetimes}
\label{sec:ffe}

The environment around astrophysical BHs is populated by electromagnetic
fields and plasma that are mutually interacting and dynamically
influenced by the BH strong-gravity. The vast majority of the
astrophysical phenomena occurring in the proximity of a BH is therefore
captured by the conservation laws for the total energy-momentum tensor
and the (conserved) currents of the system, namely,
\begin{align}
  \label{eq:conservation}
  &\nabla_{\mu} (T^{\mu\nu}_{\rm mat} + T^{\mu\nu}_{\rm em})=0\,, \\
&\nabla_{\mu} J^{\mu}_i =0\,, 
\end{align}
where $\boldsymbol{T}_{\rm mat}$ and $\boldsymbol{T}_{\rm em}$ refer
respectively to the energy-momentum content of the matter and of the
electromagnetic fields, while $\boldsymbol{J}_i$ are the conserved
currents of the charge carriers (hereafter, we will concentrate on a
single species of charge carriers, so that $\boldsymbol{J}_i \to
\boldsymbol{J}$). The symbol $\nabla$ appearing in the conservation
equations~\eqref{eq:conservation} refers to the covariant derivative with
respect to the background curved spacetime, which is employed to express
the Maxwell equations
\begin{align}
  \label{eq:maxwell}
  &\nabla_{\mu} F^{\mu\nu}=-J^\nu \,,\\
  &\nabla_{[\rho} F_{\mu\nu]}=0 \,, 
\end{align}
where $\boldsymbol{F}$ is Faraday tensor and the square brackets
indicate antisymmetric indices.
 
In a BH magnetosphere the energy and momentum of the magnetic fields are
so intense that they dominate the dynamics over the plasma, \ie
$T^{\mu\nu}_{\rm em} \gg T^{\mu\nu}_{\rm mat}$, so that $\nabla_\mu
(T^{\mu\nu}_{\rm mat} + T^{\mu\nu}_{\rm em}) \approx \nabla_\mu
T^{\mu\nu}_{\rm em} \approx 0$ and the density of the Lorentz-force can
be neglected, \ie $\nabla_\nu T^{\mu\nu}_{\rm em}=-{F^\mu}_\nu
J^\nu\approx0$. These \emph{force-free} (FF) conditions imply that a BH
magnetosphere is captured by a nonlinear regime of Maxwell's
electrodynamics called \emph{force-free electrodynamics}
(FFE)~\cite{Blandford1977, MacDonald1982, Uchida1997a, Uchida1997b}, in
which the plasma becomes a secondary quantity in the dynamical
description~\cite{Beskin1997}.

The equations of motion of FFE follow by combining the Maxwell equations
with the FF constraint
\begin{align}
  \label{eq:FFE}
  & F_{\mu\nu} \nabla_{\rho}F^{\rho\nu}=0\,, \\
  \label{eq:FFE_2}
  & \nabla_{[\rho} F_{\mu\nu]}=0\,.
\end{align}
It is important to remark that in the FF regime, $J^\mu\neq 0$,
meaning this regime is distinct from that of electrovacuum and that
matter, even though dynamically passive, is still fundamental to
sustain the magnetic fields and screen the longitudinal component of
the electric fields. In this sense, the well-known Wald
solution~\cite{Wald:74bh} is an electrovacuum solution and is not a FF
field (see, \eg \cite{Alic:2012} for a number of electrovacuum and FF
solutions).

Upon assuming that the electromagnetic field preserves the
stationarity and axisymmetry of the background spacetime, the Faraday
tensor can be rewritten as a degenerate $2$-form~\cite{Uchida1997b,
  Gralla2014} 
\begin{equation}
  \label{eq:Faraday}
  \boldsymbol{F} := d\Psi(r, \theta)\wedge \eta-\frac{\Sigma R_{_{\rm B}} }{N^2
    \sin\theta}I(r, \theta)~dr\wedge d\theta\, ,
\end{equation}
where the operator $d$ identifies the exterior derivative and $\wedge$
the antisymmetric exterior product, while the differential form
$\boldsymbol{\eta}$ is defined as
\begin{equation}
  \boldsymbol{\eta} := d\phi-\Omega_f(\Psi) dt-\frac{R_{_{\rm M}} R_{_{\rm B}}}{r
    N^2}\left[\omega-\Omega_f(\Psi)\right] dr\, .
\end{equation}
Equation~\eqref{eq:Faraday} highlights that the Faraday tensor can be
written in terms of a set of ``magnetospheric field variables'',
namely, via the ``magnetic-flux function'' $\Psi$, which is constant
along the poloidal projection of the magnetic-field component, via the
angular velocity of the magnetic-field lines $\Omega_f$, which is
related to the electric fields, and via the poloidal current $I$,
which is proportional to the toroidal magnetic field. Solving the
magnetospheric problem amounts therefore to explicitly derive
functional forms for the three quantities $\Psi$, $\Omega_f$ and $I$.
Under stationarity and axisymmetry, the FFE Eq.s~\eqref{eq:FFE} and
\eqref{eq:FFE_2} can be combined in the \emph{Grad-Shafranov} (GS)
equation for BH magnetospheres [see Eq.~\eqref{eq:GSE}], and the
magnetospheric field variables can be derived by imposing appropriate
regularity conditions. We refer the reader to App.~\ref{app:BHMAG} for
a precise formulation of the properties of FFE magnetospheres in KRZ
spacetimes via the GS equation.\\

\subsection{BZ jet luminosity in KRZ spacetimes}
\label{sec:BZJET}

As mentioned in Sec.~\ref{sec:intro}, the BZ
mechanism~\cite{Blandford1977}, which can be interpreted as a purely
electrodynamical manifestation of the Penrose process~\cite{Lasota2014},
extracts energy and angular momentum from a spinning BH in the presence
of a stationary and axisymmetric FF magnetosphere. This mechanism is
currently considered the best theoretical candidate to explain how
relativistic jets are powered and sustained in active galactic nuclei and
short gamma-ray bursts.

The derivation of the power extracted in this process, \ie the BZ
luminosity $P_{_{\rm BZ}}$, closely follows the standard derivation in
GR. In the interest of not detracting the reader's attention with
technical aspects, we summarise here the main step of the derivation
and refer the reader to App.~\ref{sec:ffepert}, \ref{app:BZ} and
\ref{app:shooting} for more detailed discussions.

In our approach, we first perturbatively expand the field variables,
assuming that for small BH spin, $a_*\ll1 $, a FF solution will be a
correction to an electrovacuum field with magnetic flux $\psi_0$,
hence $\Psi\sim\psi_0+ \mathcal{O}(a_*)$, $I\sim\mathcal{O}(a_\star)$,
$\Omega_f\sim\mathcal{O}(a_*)$. We then explicitly solve the BH
magnetospheric problem order-by-order in the perturbation theory. Once
a solution is known to a given perturbative order, the flux of
electromagnetic energy-momentum tensor can be explicitly computed, so
as to obtain the BZ luminosity as a function of the BH angular
velocity $P_{_{\rm BZ}}(\Omega_h)$ [see Eq.~\eqref{eq:BZ1}].

In the specific case of a \emph{split-monopole magnetosphere} around a
generic KRZ BH, which exists regardless of the choice of the KRZ
parameters (see App.~\ref{sec:ffepert}), a simple power-counting via
Eq.~\eqref{eq:BZ} reveals that the perturbative order we resolved is
sufficient to express the BZ power in KRZ, \ie $(P_{_{\rm BZ}})_{_{\rm
    KRZ}}$, up to the 6th order in the BH angular velocity
$\Omega_h$. The corresponding theory-agnostic expression for the BZ
luminosity in the KRZ background then reads
\begin{widetext}
\begin{align}
  \label{eq:BZKRZ}
  (P_{_{\rm BZ}})_{_{\rm KRZ}} = &\kappa \, (2\pi\Psi_h)^2 \,
  \Omega_h^2 \, f_{_{\rm KRZ}} (\Omega_h) =
  \frac{2}{3}\pi\Psi_h^2\Omega_h^2\bigg\{1 +
  \frac{16}{5}\Big[(1+\varrho)^2-4\mathcal{R}_{22}^{h}\Big]
  \frac{M^2\Omega_h^2}{(1+\varrho)^4} \nonumber \\
  &+\frac{256}{5}\left[{\mathcal{R'}}_{22}^{h}-4\mathcal{R}_{42}^{h} +
    \frac{9+14\mathcal{R}_{22}^{h}-19\varrho
      (2+\varrho)}{7}\mathcal{R}_{22}^{h} +
    (1+\varrho)^2\frac{23\varrho(2+\varrho) -
      5}{56}\right]\frac{M^4\Omega_h^4}{(1+\varrho)^8}\bigg\}
  +\mathcal{O}(\Omega_h^7)\,,
\end{align}
where $\Psi_h$ is the magnetic flux at the horizon and the quantities
$\mathcal{R}^h_{22}$, $\mathcal{R}_{42}^h$, and
${\mathcal{R}'}^h_{22}$ (where ${\mathcal{R}'}^h_{mn} :=
(d\mathcal{R}_{mn}/dr)_{r_0}$) are related to the horizon value of the
magnetic flux expansion [see Eq.~\eqref{eq:Psi}].
\end{widetext}

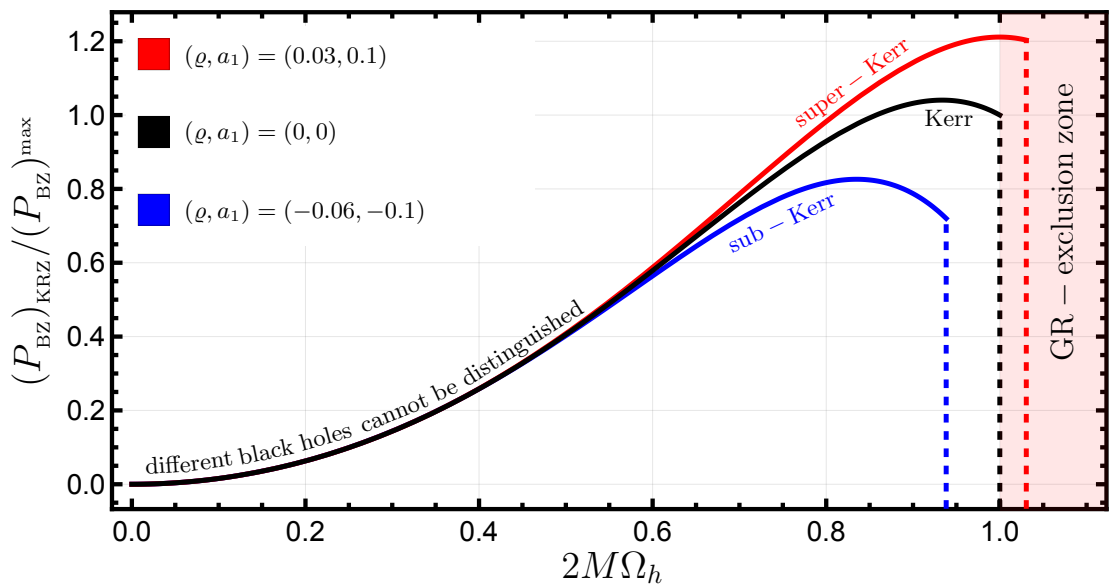
\begin{figure*}[ht]
  \centering
  \includegraphics[width=0.8\textwidth]{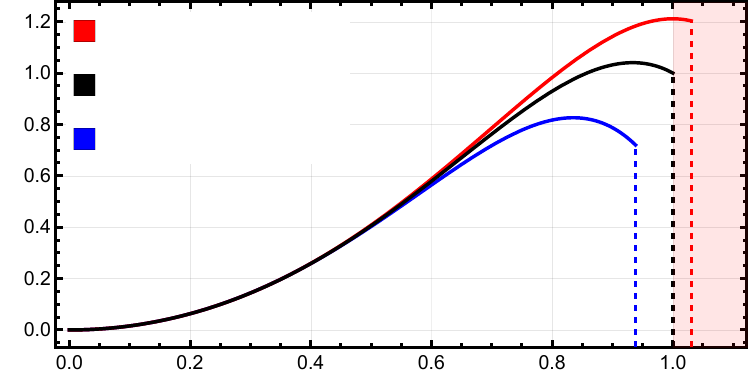}
  \begin{picture}(0, 0)
    \put(-355, 183){ $(\varrho, a_1)=(0.03, 0.1)$} \put(-355, 124){
      $(\varrho, a_1)=(-0.06, -0.1)$} \put(-355, 154){
      $(\varrho, a_1)=(0, 0)$ }
    \put(-370, 27){\color{black}{\rotatebox{4}{ ${\rm diff}$}}}
    \put(-357, 27.5){\color{black}{\rotatebox{8}{ ${\rm erent~}$}}}
    \put(-335, 30.3){\color{black}{\rotatebox{11.5}{ ${\rm black~}$}}}
    \put(-312, 35.5){\color{black}{\rotatebox{15.5}{ ${\rm holes}$}}}
    \put(-289, 42){\color{black}{\rotatebox{22.5}{ ${\rm cannot }$}}}
    \put(-262, 53.5){\color{black}{\rotatebox{25}{ ${\rm be }$}}}
    \put(-251, 59){\color{black}{\rotatebox{32.5}{ ${\rm distinguished }$}}}
    \put(-125, 157){\color{red}{\rotatebox{36}{ ${\rm super-Kerr}$}}}
    \put(-75, 159){\color{black}{ ${\rm Kerr}$}}
    \put(-150, 110){\color{blue}{\rotatebox{25}{ ${\rm sub-Kerr}$}}}
    \put(-210, -12){\Large $2M\Omega_h$}
    \put(-418, 60){\rotatebox{90}{\Large $(P_{_{\rm BZ}})_{_{\rm
            KRZ}}/(P_{_{\rm BZ}})^{_{\rm max}}$}}
    \put(-25, 70){\color{black}{\rotatebox{90}{\large ${\rm GR-exclusion
          ~zone}$}}}
   \end{picture}
  \vspace{2mm}
  \caption{Variation of the BZ luminosity in a generic KRZ spacetime
    $(P_{_{\rm BZ}})_{_{\rm KRZ}}$ [Eq.~\eqref{eq:BZKRZ}], when
    normalised by the power extracted by an extreme Kerr BH $(P_{_{\rm
        BZ}})^{_{\rm max}}$ [Eqs.~\eqref{eq:BZ2} and \eqref{eq:f_GR}],
    shown as a function of the (dimensionless) BH angular velocity
    $2M\,\Omega_h$. Reported with a black solid line is the power in a
    Kerr spacetime, while solid red and blue lines highlight two
    representative choices of the parameters $\varrho$ and $a_1$ that
    lead to super-Kerr and sub-Kerr behaviours, respectively. For these
    spacetimes, the relative variations in the Kretschmann scalar are
    $\Delta_\% \approx 41\%$ (super-Kerr) and $\Delta_\% \approx- 41\%$
    (sub-Kerr). The curves are truncated at the maximum value of
    $\Omega_h$ consistent with the constraints on the KRZ parameters and
    the red-shaded are refers to solutions not allowed in GR. Note that
    although slowly rotating BHs can hardly be distinguished, significant
    changes appear for $\Omega_h \gtrsim 0.3/M$.}
\label{fig:BZ}
\end{figure*}

A number of considerations are now in order to appreciate the
consequences of Eq.\eqref{eq:BZKRZ}.
\begin{itemize}
\item Despite the complexity of the KRZ metric, the BZ luminosity in a
  generic KRZ spacetime is remarkably compact. More importantly,
  comparing expression~\eqref{eq:BZKRZ} with the corresponding one in
  GR, highlights the universality of the \emph{functional form} of the
  BZ power emitted from the (split-monopole) magnetosphere of
  stationary and axisymmetric BHs, namely $\kappa \, (2\pi\Psi_h)^2 \,
  \Omega_h^2 \, f_{_{\rm KRZ}} (\Omega_h)$.
  Moreover, the expression above is valid for every choice of the KRZ
  parameters.  While we have demonstrated this at
  $\mathcal{O}(\Omega_h^6)$, there is no obvious mathematical reason
  why it would not hold at higher orders.
\item The global prefactor $\kappa$ referring to the magnetic-field
  topology~\cite{Meringolo2025a} matches exactly the value for a split
  monopole in GR, \ie \hbox{$\kappa = 1/6\pi$}.
\item The lowest-order contribution in \eqref{eq:BZKRZ} reflects the
  quadratic scaling $(P_{_{\rm BZ}})_{_{\rm KRZ}}\sim\Omega_h^2$ obtained
  by Blandford and Znajek~\cite{Blandford1977} and highlights the
  \emph{universality of the BZ power in slowly rotating BH spacetimes}.
   This degeneracy for $a_* \ll 1$ points to an effective difficulty in
  distinguishing different slowly rotating BHs by measuring the power of
  the corresponding relativistic jets.
\item The next-to-leading-order contributions are encoded in the
  high-spin factor $f_{_{\rm KRZ}}(\Omega_h)$, which depends on the
  entire set of the KRZ parameters. Besides $\varrho$, which appears
  explicitly in Eq.~\eqref{eq:BZKRZ}, all the KRZ parameters enter via
  the terms ${\mathcal{R}}_{22}^{h}$, $\mathcal{R}_{42}^{h}$, and
  ${\mathcal{R'}}_{22}^{h}$. Their expression as function of the KRZ
  parameters $(\varrho, a_1, b_1)$ is computed explicitly in
  App.~\ref{app:shooting} and are used in the next section.
\item When the KRZ parameters vanish, that is, when the KRZ spacetime
  reduces to the Kerr solution, the terms ${\mathcal{R}}_{22}^{h}$,
  $\mathcal{R}_{42}^{h}$, and ${\mathcal{R'}}_{22}^{h}$ assume their
  general-relativistic expressions, \ie ${\mathcal{R}}_{22}^{h} =
  {(6\pi^2 - 49)}/{72}$, ${\mathcal{R'}}_{22}^{h} = {(6\pi^2 - 61)}/{12}
  $, and $\mathcal{R}_{42}^{h} = {39\zeta(3)}/{3920} +
  {17929399}/{2540160} - {3877\pi^2}/{12096} - {19\pi^4}/{480}$, where
  $\zeta(n)$ is the Riemann zeta function.
\item The non-trivial dependence of $f_{_{\rm KRZ}}(\Omega_h)$ on the
  spacetime background breaks the aforementioned
  degeneracy. Therefore, the BZ luminosity of rapidly rotating BHs can
  be to used to discriminate among different spacetimes if independent
  measurements of the BH spin are available.
\item Notwithstanding the perturbative nature of this result, the
  inclusion of $\mathcal{O}(\Omega^6_h)$ corrections have been shown
  to be necessary in order to make the quantitative estimates accurate
  enough also in the regime of rapid rotation, as demonstrated in
  Refs.~\cite{Camilloni2022}. The inclusion of even higher orders in
  the perturbative expansion and the comparison with numerical
  simulations of the BZ power around KRZ black holes, will further
  strengthen the results of our analysis also in the regime $a_* \to
  1$~\cite{Meringolo2025a}. At the same time, the important message we
  provide with our paper about the universality of the BZ power at
  slow rotation rates is mathematically very robust as it applies to
  regimes where the perturbative approach is most accurate.
\end{itemize}

\subsection{Comparison with Kerr spacetimes}
\label{sec:BZ_plots}

\begin{figure}[t]
  \includegraphics[width=0.45\textwidth]{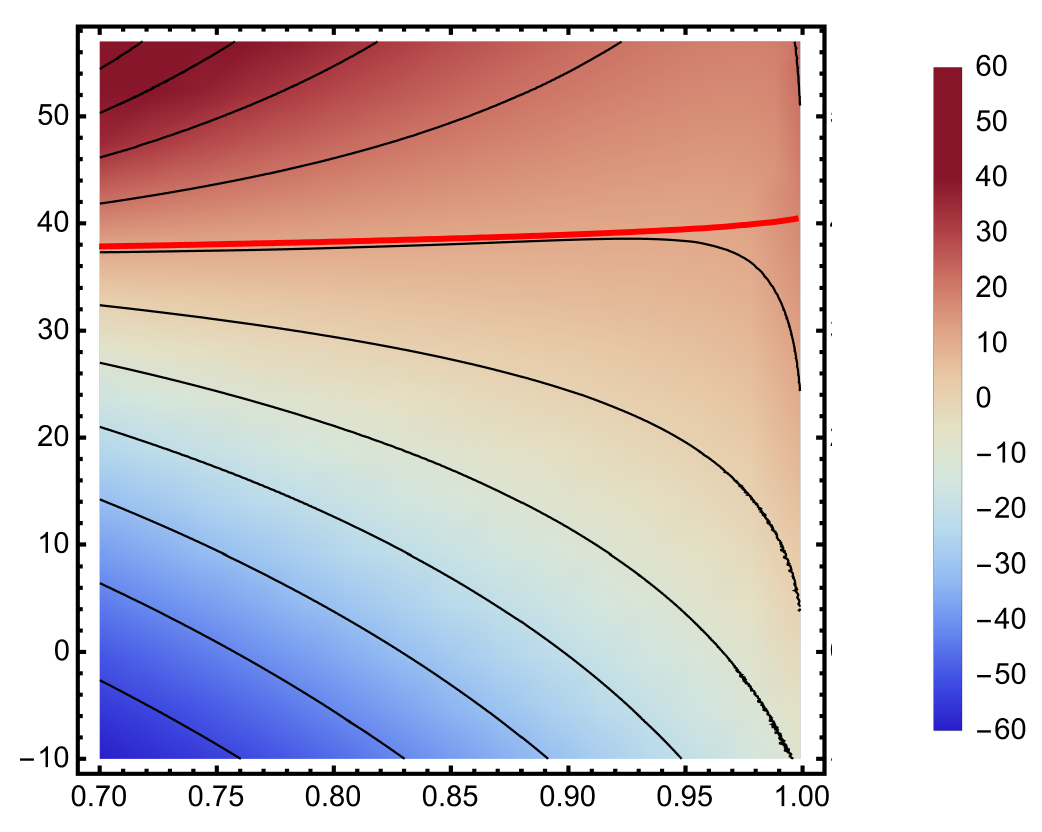}
  \begin{picture}(0, 0)
     \put(-147, 182){\footnotesize{$a_1=0.1$}}
     \put(-240, 90){\rotatebox{90}{\large$\Delta_\%$}}
     \put(-135, -8){\large$a_*$}
     \put(-120, 135){\color{red}{ $\varrho=0.03$}}
     \put(-208, 26){\rotatebox{-30}{\scriptsize$-50$}}
     \put(-190, 38){\rotatebox{-27}{\scriptsize$-40$}}
     \put(-172, 49){\rotatebox{-30}{\scriptsize$-30$}}
     \put(-155, 61){\rotatebox{-30}{\scriptsize$-20$}}
     \put(-135, 73){\rotatebox{-27}{\scriptsize$-10$}}
     \put(-105, 90){\rotatebox{-30}{\scriptsize$0$}}
     \put(-75, 124){\rotatebox{-35}{\scriptsize$10$}}
     \put(-130, 151){\rotatebox{20}{\scriptsize$20$}}
     \put(-172, 156){\rotatebox{28}{\scriptsize$30$}}
     \put(-193, 161){\rotatebox{30}{\scriptsize$40$}}
     \put(-208, 164){\rotatebox{30}{\scriptsize$50$}}
     \put(-42, 24){\rotatebox{90}{\normalsize $\Delta P_{_{\rm BZ}} := (P_{_{\rm BZ}})_{_{\rm KRZ}}/(P_{_{\rm BZ}}) - 1~[\%]$}}
   \end{picture}
   \\
   \vspace{8mm}
   \includegraphics[width=0.45\textwidth]{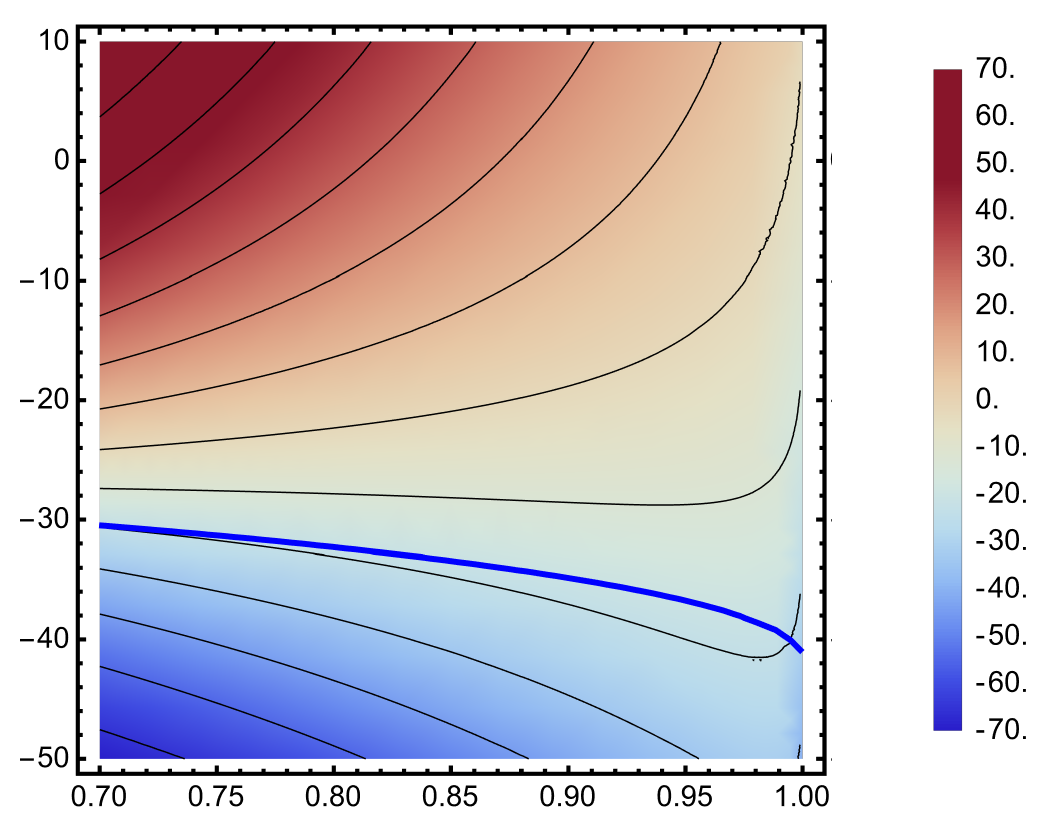}
   \begin{picture}(0, 0)
     \put(-146, 182){\footnotesize{$a_1=-0.1$}}
     \put(-240, 90){\rotatebox{90}{\large$\Delta_\%$}}
     \put(-135, -8){\large$a_*$}
     \put(-120, 62){\rotatebox{-10}{\color{blue}{ $\varrho=-0.06$}}}
     \put(-205, 29){\rotatebox{-20}{\scriptsize$-50$}}
     \put(-190, 38){\rotatebox{-20}{\scriptsize$-40$}}
     \put(-172, 45){\rotatebox{-15}{\scriptsize$-30$}}
     \put(-155, 55){\rotatebox{-13}{\scriptsize$-20$}}
     \put(-135, 76){\rotatebox{0}{\scriptsize$-10$}}
     \put(-105, 102){\rotatebox{15}{\scriptsize$0$}}
     \put(-138, 115){\rotatebox{25}{\scriptsize$10$}}
     \put(-157, 125){\rotatebox{35}{\scriptsize$20$}}
     \put(-172, 135){\rotatebox{35}{\scriptsize$30$}}
     \put(-185, 145){\rotatebox{35}{\scriptsize$40$}}
     \put(-196, 154){\rotatebox{40}{\scriptsize$50$}}
     \put(-208, 164){\rotatebox{40}{\scriptsize$60$}}
     \put(-42, 24){\rotatebox{90}{\normalsize $\Delta P_{_{\rm BZ}} := (P_{_{\rm BZ}})_{_{\rm KRZ}}/(P_{_{\rm BZ}}) - 1~[\%]$}}
   \end{picture}
   \\
   \vspace{3mm}
  \caption{Colormap showing the relative deviation between the BZ power
    emitted from a KRZ BH and that from a Kerr BH, $\Delta P_{_{\rm BZ}}$
    shown as a function of the BH spin $a_*$ and of the fractional
    spacetime deviation from Kerr $\Delta_\%(\varrho,a_1)$. The latter
    depends on the KRZ parameters $\varrho$ and $a_1$ ($b_1=0$), and is
    presented for $a_1=0.1$ (top panel) and for $a_1=-0.1$ (bottom
    panel). The solid red and blue lines represent curves of constant
    $\varrho$ associated to a super-Kerr BH ($\varrho=0.03$) and to a
    sub-Kerr one ($\varrho=-0.06$), matching the representative cases
    shown in Fig.~\ref{fig:BZ}.}
\label{fig:BZSTcorrelation}
\end{figure}
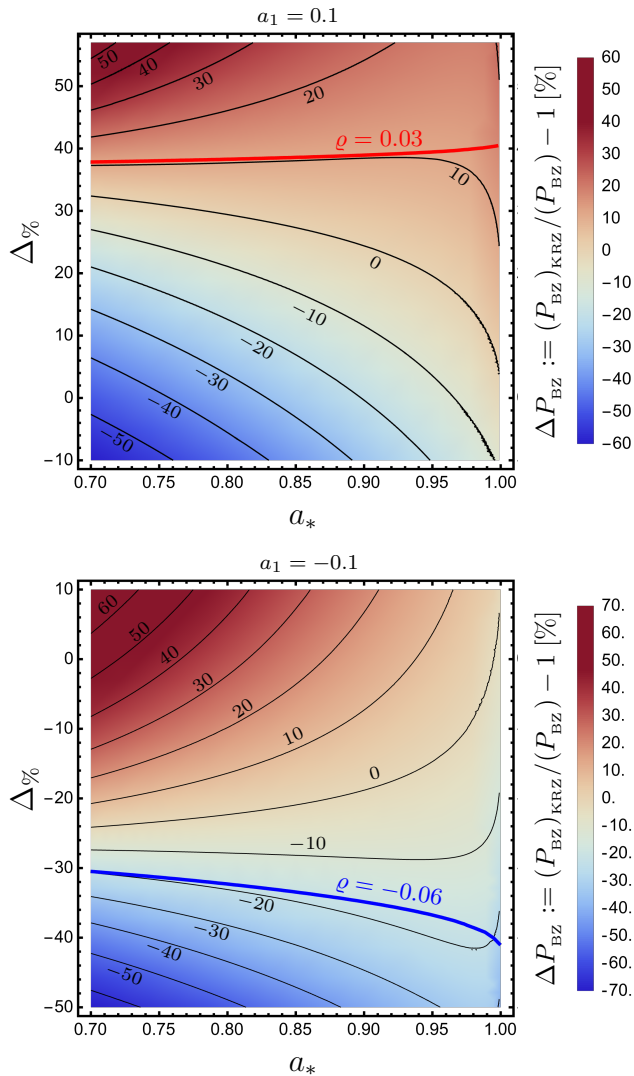

In this section we provide explicit estimates for the rate of energy
extracted via the BZ mechanism in a KRZ spacetime, $(P_{_{\rm
    BZ}})_{_{\rm KRZ}}$, given in Eq.~\eqref{eq:BZKRZ}. However, before
turning to the actual results on $(P_{_{\rm BZ}})_{_{\rm KRZ}}$, a
technical note should be made on its numerical calculation, which
involves the determination of the quantities ${\mathcal{R}}_{22}^{h}$,
$\mathcal{R}_{42}^{h}$ and ${\mathcal{R'}}_{22}^{h}$. An expression for
the latter can be algebraically derived from the near-horizon Frobenius
series for the function $\mathcal{R}_{22}(r)$ (see Eq.~\eqref{eq:
  exp1}) and reads (generalising this expression for arbitrary choices of
the KRZ parameters is straightforward)
\begin{equation}
  \label{eq:DR22H}
  {\mathcal{R'}}_{22}^{h} := \frac{(1+b_1)^2\left[6{\mathcal{R}}_{22}^{h}
      (1+a_1-2\varrho) - (1+\varrho)^2\right]}{(1-a_1-2\varrho)^2}\, .
\end{equation}
Using this relation, the calculation of the BZ power ultimately amounts
to computing the horizon values ${\mathcal{R}}_{22}^{h}$ and
${\mathcal{R}}_{42}^{h}$. This requires solving the ordinary differential
equations for the radial magnetic-flux functions $\mathcal{R}_{22}(r)$
and $\mathcal{R}_{42}(r)$ by combining the Frobenius expansions with a
shooting method (see App.~\ref{app:shooting} for details). In this way,
we are able to cover efficiently the whole space of parameters $(a_*,
\varrho, a_1, b_1)$ while ensuring that for vanishing KRZ parameters we
recover the values expected 6th-order analytic result for the BZ power in
the Kerr spacetime, \ie ${\mathcal{R}}_{22}^{h}\approx0.142$ and
${\mathcal{R}}_{42}^{h}\approx0.051$.

Figure~\ref{fig:BZ} offers a comparison of the BZ luminosity in sub-Kerr
and super-Kerr spacetimes by combining the KRZ parameters $\varrho$ and
$a_1$ in such a way as to attain a comparable spacetime fractional
deviation between the two cases, with $|\Delta_\%|\approx 41\%$ (we here
keep $b_1 = 0$ as this parameter has only marginal impact on the BZ
luminosity; see legend for details). The power is normalized to the power
extracted by an extreme Kerr BH, $(P_{_{\rm BZ}})^{_{\rm max}} \approx
0.337/M^2$, with $\Omega_h=1/2M$. In this way, it becomes apparent how
even largely different but slowly-rotating BH spacetimes are virtually
indistinguishable in terms of the corresponding BZ power. More
importantly, different behaviours in $(P_{_{\rm BZ}})_{_{\rm KRZ}}$ start
to appear only at large horizon angular velocities, \ie for
$\Omega_h\gtrsim 0.3/M$. Fortunately, for larger spins, differences of
$\simeq 20-30\%$ appear in the BZ luminosity, so that a relativistic jet
that is brighter or dimmer than what expected in GR can be used as a
probe of the spacetime properties.

Note that the BZ luminosity curves in Fig.~\ref{fig:BZ} do not show a
monotonic dependence on $\Omega_h$. This is an artefact of the
perturbative expansion that disappears by including higher-order
corrections, as shown in the Kerr case~\cite{Camilloni2022}. Furthermore,
the curves terminate at different values of the maximum BH horizon
angular velocity as a result of the physical constraints on the KRZ
parameter space and of the sub/super-Kerr nature of the BHs considered
(see Sec.~\ref{sec:KRZ}). Indeed, while $\Omega_h = 1/2 M$ corresponds to
the maximum angular velocity attained for a Kerr BH at extremality ($a_*
= 1$), super-Kerr BHs can exceed this limiting value and extract energy
via the BZ mechanism even for $2M\,\Omega_h > 1$, in what we mark as
``GR-exclusion zone'' in Fig.~\ref{fig:BZ}\footnote{A similar behaviour
has been recently observed in via GRMHD simulations of accreting black
holes in parameterised metrics~\cite{Chatterjee2023}.}.

So far, different BHs have been compared in terms of the luminosity of
their jets emitted at the same angular velocity $\Omega_h$. It
nonetheless also instructive to also draw comparisons for the same BH
spin $a_*$. While in GR the two quantities are related via a nonlinear
expression, \ie \hbox{$2M\,\Omega_h = a_*/[M+\sqrt{M^2-a^2_*}]$}, as
mentioned in Sec.~\ref{sec:subsuperKerr}, in alternative theories of
gravity this is not so obvious, and additional parameters can mix in a
degenerate manner. This is indeed a feature well illustrated by the KRZ
metric, where $\Omega_h(a_*, \varrho)$ [see Eq.~\eqref{eq:omegaH} and
  Fig.~\ref{fig:omega}].

Also interesting is to estimate to which extent the deviations from GR,
quantified by $\Delta_\%$, reflect into deviations in the BH jet
luminosity from the corresponding Kerr case. We do this in
Fig.~\ref{fig:BZSTcorrelation}, which illustrates how the percentage
deviation of the BZ power with respect to the Kerr case correlates with
the BH spin $a_*$ and the spacetime deviation $\Delta_\%(a_*, \varrho,
a_1)$ produced by the KRZ parameters. To make contact with the example
previously considered, the two panels in Fig.~\ref{fig:BZSTcorrelation}
refer to $a_1=\pm 1$ and coloured thick curves represent the
\hbox{$\varrho=0.03$} and \hbox{$\varrho=-0.06$} constant lines with the
same colours used in Fig.~\ref{fig:BZ}. Considering $a_*=0.95$ as a
representative value for the BH spin, we can estimate how the BZ jet
power deviates from the theoretical prediction in GR: in this case, the
jet launched by a super-Kerr BH with $(\varrho, a_1)=(0.03, 0.1)$, that
is, a KRZ BH with a $\Delta_\%\approx40\%$ deviation from GR, would
launch a jet $\approx10\%$ brighter than the jet emitted by an equally
spinning Kerr BH; conversely, for a sub-Kerr BH with $(\varrho,
a_1)=(-0.06, -0.1)$, with KRZ deviation $\Delta_\%\approx-37\%$, the jet
emitted would be $\approx 20\%$ dimmer than the jet emitted in the GR
case.

\begin{figure*}[ht]
  \centering
  \includegraphics[width=.9\textwidth]{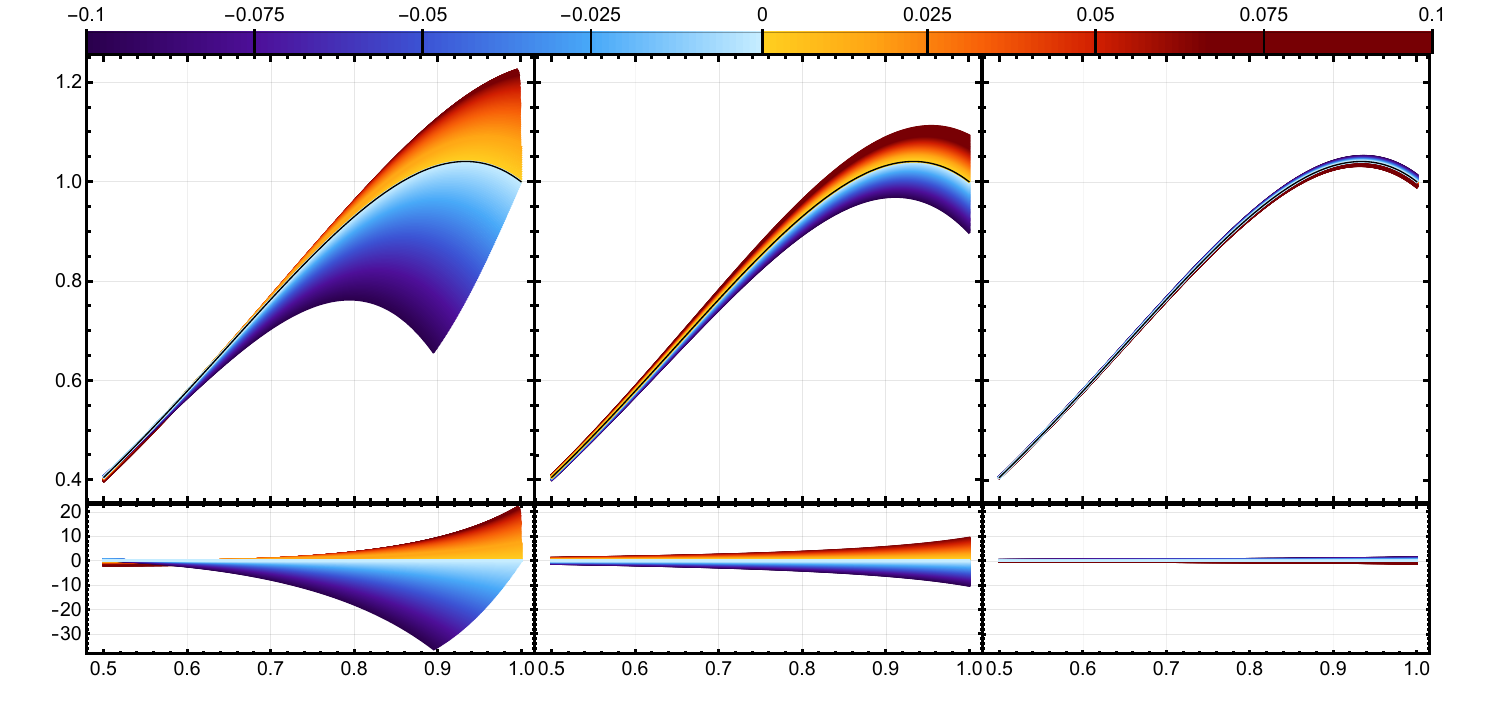}
  \begin{picture}(0, 0)
    \put(-420, 180){ only $\varrho$ varied}
    \put(-280, 180){ only $a_1$ varied}
    \put(-150, 180){ only $b_1$ varied} \put(-245, 220){ $\varrho, a_1, b_1$}
    \put(-375, -5){ $2M\Omega_h$} \put(-240, -5){ $2M\Omega_h$}
    \put(-105, -5){ $2M\Omega_h$} \put(-460, 100){\rotatebox{90}{\large
        $(P_{_{\rm BZ}})_{_{\rm KRZ}}/(P_{_{\rm BZ}})^{_{\rm
            max}}$}}
    \put(-455, 20){\rotatebox{90}{
        \footnotesize$\Delta P_{_{\rm BZ}}~[\%]$}}
   \end{picture}
  \vspace{2mm}
  \caption{\textit{Top panels:} as in Fig.~\ref{fig:BZ}, we show the
    variation of the BZ luminosity in a generic KRZ spacetime
    normalised to that of an extreme Kerr BH as a function of the
    (dimensionless) BH angular velocity. The different panels are used
    to illustrate the dependence on the three-dimensional space of KRZ
    parameters $(\varrho, a_1, b_1)$; in particular, from left to
    right, the three panels show the variations when only one of such
    parameters is varied in the interval $[-0.1, 0.1]$ while the other
    two are set to zero. The colormap reported above provides a
    measure of the excursion of the varied parameter. \textit{Bottom
      panels} the percentage deviation in the BZ luminosity for the
    various cases considered in the top panels.}
\label{fig:BZcat}
\end{figure*}
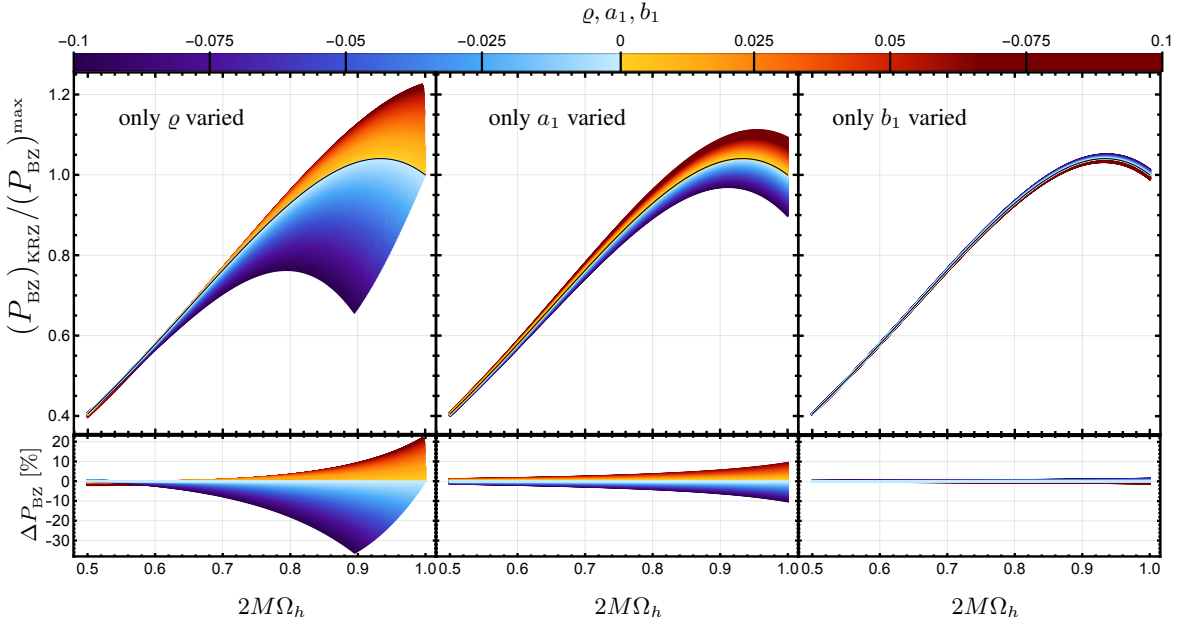
%

\subsection{Exploring the space of parameters of BZ luminosities}
We have seen in the previous section that the general functional relation
for the luminosity of a relativistic jet launched via the BZ mechanism in
terms of the BH angular velocity is given by Eq.~\eqref{eq:BZKRZ} and can
be schematically be written as \hbox{$(P_{_{\rm BZ}})_{_{\rm KRZ}}\sim
  \Omega_h^2 \, f_{_{\rm KRZ}}(\Omega_h)$}, where $f_{_{\rm
    KRZ}}(\Omega_h) \neq 1$ only in the high-spin
regime. Equation~\eqref{eq:BZKRZ} should be seen as a function depending
on a four-dimensional space of parameters given by $(a_*, \varrho, a_1,
b_1)$. Clearly, exploring such a large space is difficult but essential
in order to assess how different deviations from GR can affect the BZ
luminosity. Hence, hereafter we will concentrate on sections of this
space of parameters where one or two of such parameters are being held
fixed or set to zero.

However, even when considering cross-sections of the space of parameters,
we find it important (and computationally less expensive) to be able to
write the function $f_{_{\rm KRZ}}(\Omega_h)$ in terms of simple analytic
expressions that capture its behaviour when only one of the KRZ
parameters is switched on and the other ones are set to zero. This is
possible because the quantities ${\mathcal{R}}_{22}^{h}$ and
${\mathcal{R}}_{42}^{h}$ -- that are ultimately needed to compute
$f_{_{\rm KRZ}}(\Omega_h)$ -- can be well approximated with quadratic
polynomials when the KRZ parameters are in the interval $\varrho, a_1,
b_1 \in [-0.1, 0.1]$ [see Eq.~\eqref{eq:R22R42appr} in
  Appendix~\ref{app:shooting} for the explicit expressions]. Such
polynomial approximations for ${\mathcal{R}}_{22}^{h}$ and
${\mathcal{R}}_{42}^{h}$ allow us to obtain analytic expressions for the
spin-induced corrections to the BZ luminosities of the type
\begin{align}
  \label{eq:f123_a_old}
  f^{\varrho}_{_{\rm KRZ}}(\Omega_h)&\approx 1+\left(1.38-7.42\varrho+5.9
  \varrho^2\right)M^2\Omega_h^2
  \\\nonumber
  &-\left(11.25 -107.3\varrho
  +404.7\varrho^2\right)M^4 \Omega_h^4
  +\mathcal{O}\left(\Omega_h^6, \varrho^3\right)\, ,
  \\\nonumber
  \\
  \label{eq:f123_b_old}
  f^{a_1}_{_{\rm KRZ}}(\Omega_h)&\approx 1+\left(1.38+1.72a_1-1.62
  a_1^2\right)M^2\Omega_h^2
  \\\nonumber
  &-\left(11.25 -1.67a_1 -4.68a_1^2\right)M^4
  \Omega_h^4 +\mathcal{O}\left(\Omega_h^6, a_1^3\right)\, ,
  \\\nonumber
  \\
  \label{eq:f123_c_old}
  f^{b_1}_{_{\rm KRZ}}(\Omega_h)&\approx 1+\left(1.38-0.41b_1-0.34
  b_1^2\right)M^2\Omega_h^2
  \\\nonumber
  &-\left(11.25 -0.46b_1 -0.2b_1^2\right)M^4
  \Omega_h^4 +\mathcal{O}\left(\Omega_h^6, b_1^3\right)\, .
\end{align}
where, \eg $f^{\varrho}_{_{\rm KRZ}}(\Omega_h) := f_{_{\rm
    KRZ}}(\Omega_h, {\varrho})$ with $a_1=0=b_1$.
\begin{widetext}
Because of their
systematic functional form,
Eqs.~\eqref{eq:f123_a_old}--\eqref{eq:f123_c_old} can be also cast in a
more compact form as

\begin{equation}
  \label{eq:f123}
  f^{Q_{(j)}}_{_{\rm KRZ}}(\Omega_h)= 1+\left(1.38+\sum_{i=1}^{2}\alpha_{(j), n} 
  (Q_{(j)})^n \right)M^2\Omega_h^2+\left(-11.25+\sum_{i=1}^{2}\beta_{(j), n}
  (Q_{(j)})^n \right)M^4\Omega_h^4+\mathcal{O}\left(\Omega_h^6, Q^3_{(j)}\right)
\end{equation}
where $Q_{(j)}=\left\{ \varrho, a_1, b_1 \right\}$
\begin{align}
  &\alpha_{(1)} = \left\{ -7.42, 5.90 \right\}\, , & 
  &\alpha_{(2)} = \left\{ 1.72, -1.62 \right\}\, , & 
  &\alpha_{(3)} = \left\{ -0.41, -0.34 \right\}\, ,  \\
  &\beta_{(1)} = \left\{ -107.30, 404.70 \right\}\, , &
  &\beta_{(2)} = \left\{ -1.67, -4.68 \right\}\, , & 
  &\beta_{(3)} = \left\{ -0.46, -0.20 \right\}\, . 
\end{align}
\end{widetext}

The importance of Eqs.~\eqref{eq:f123_a_old}--\eqref{eq:f123_c_old} --
which provide analytic estimates that case differ from the numerical ones
at most by $5\%$ -- is that they represent the first theory-agnostic
description of the BZ luminosity in generic BH spacetimes captured by
two-dimensional sections of the four-dimensional space of solutions
spanned by the parameters $\Omega, \varrho, a_1, b_1$. Being analytic,
expression~\eqref{eq:f123_a_old}--\eqref{eq:f123_c_old} provide immediate
coverage of the behaviour of $(P_{_{\rm BZ}})_{_{\rm KRZ}}$ in three
relevant sections. This is shown in Fig.~\ref{fig:BZcat}, whose panels
show the variance of $(P_{_{\rm BZ}})_{_{\rm KRZ}}$ as a function of
$\varrho$ (left panel), $a_1$ (middle panel), and $b_1$ (right panel),
when normalised to the corresponding value in GR, \ie $P_{_{\rm BZ}}$.

In this way, it is possible to appreciate that variations in $\varrho$
and $a_1$ can lead to significant deviations of the BZ power with respect
to the Kerr case already when their effect on the spacetime leads to
$|\Delta_\%|\approx 20\%$. This is to be expected, given the nature of
the BZ mechanism as a generalised Penrose process. Indeed, the energy
extraction is expected to occur between the horizon radius and the
ergosphere~\cite{Lasota2014}. While the first is directly controlled by
$\varrho$, the latter is determined by $g_{tt}=0$, in which also $a_1$
enters. The parameter $b_1$, on the other hand, never affects the region
of energy extraction and, in fact, produces only marginal deviations of
the BZ power even when its value is associated to large deviations from
the Kerr spacetime $\Delta_\%(b_1)\approx 70\%$, for $b_1=-0.2$. As a
concluding remark, we note that the chosen range $\varrho, a_1,
b_1\in[-0.1, 0.1]$ allows us to probe non-negligible degrees of spacetime
deviations $\Delta_\%$ (see Fig.~\ref{fig:delta}); such a range, however,
can be easily broadened.

\section{Conclusion}

The Blandford-Znajek (BZ) mechanism represents a cornerstone in
relativistic astrophysics and is commonly regarded as a credible
process through which rotational energy from an accreting black hole
can be extracted and used to power the emission of relativistic
jets. This electrodynamical manifestation of the Penrose process has
been explored in great detail in GR, and it is now clear that the
extracted power is a function of the BH spin and can be expressed
schematically as \hbox{$P_{_{\rm BZ}} \sim \Omega_h^2\, f(\Omega_h)$},
where the function $f(\Omega_h)$ contains high-order contributions of
the horizon angular velocity $\Omega_h$ and where $f(\Omega_h) \simeq
1$ for $M \Omega_h \ll 1$, as originally found by Blandford and
Znajek. Excluding some remarkable exceptions, the validity of the BZ
mechanism in other theories of gravity and hence in other BH spacetime
is essentially unexplored.

For this reason, we have here explored the efficiency of this process for
generic black-hole spacetimes within a well-known and versatile
parameterisation of stationary and axisymmetric spacetimes, \ie the KRZ
expansion. In this way, we have revealed a a \emph{universal} nature of
the leading-order behaviour, namely, that also in generic KRZ spacetimes,
the BZ luminosity scales as a quadratic function of the BH angular
velocity. Furthermore, we have shown that the mathematical structure of
the agnostic BZ luminosity can be cast into a generic form
\hbox{$(P_{_{\rm BZ}})_{_{\rm KRZ}} \sim \Omega_h^2\, f_{_{\rm
      KRZ}}(\Omega_h)$}, where $f_{_{\rm KRZ}}(\Omega_h)$ has been
computed up to $\mathcal{O}(\Omega^4_h)$ and it depends on the details of
the background spacetime via the coefficients of the KRZ expansion
$\varrho, a_1$ and $b_1$. Importantly, we have shown that the
spin-dependence function $f_{_{\rm KRZ}}(\Omega_h)$ can be written with a
very good approximation in terms of simple analytic expressions, thus
allowing us to explore a large space of parameters that has been so far
explored in terms of expensive numerical simulations.

One of the most important results of our analysis can be arguably be
considered the demonstration of the quasi-universal scaling of the BZ
luminosity, \ie that also $f_{_{\rm KRZ}}(\Omega_h) \simeq 1$ for $M
\Omega_h \ll 1$. This conclusion has the important implication that is
intrinsically difficult to distinguish different theories of gravity by
measuring the luminosity of jets launched from BHs that are
slowly-rotating. At the same time, the fact that $f_{_{\rm
    KRZ}}(\Omega_h) \neq 1$ for $\Omega_h \gtrsim 0.3 / M$ and the
non-trivial dependence of $f_{_{\rm KRZ}}(\Omega_h)$ on the spacetime
background, break the degeneracy among slowly rotating BH spacetimes. As
a result, the BZ luminosity of rapidly rotating BHs can be to used to
discriminate among different spacetimes. Although exciting, the use of
the BZ jet luminosity as a strong-gravity probe poses both observational
and theoretical challenges. Indeed, it requires not only high-precision
and independent measurements the BH angular velocity, but also a
high-degree of confidence on the magnetic-field topology on horizon
scales that can also impact the efficiency of the BZ emission. Clearly,
these are fascinating challenges to be tackled both observationally and
theoretically.

While deriving these results, a number of important ancillary conclusions
were also obtained. For instance, we have determined a set of constraints
for the KRZ parameter space that are a generalisation of the ones derived
for the RZ metric in spherical symmetry. Moreover, we used the
Kretschmann scalar to define a quantifier $\Delta_\%$ that provides a
percentage measure for the deviation of a KRZ BH from the Kerr
spacetime. This quantity is important to give a precise meaning to the
values assigned to the KRZ parameters in terms of the extent to which
they affect the spacetime.

Although this is the first analytic extension of the calculation of the
BZ mechanism in generic stationary and axisymmetric BH spacetimes, also
this work is subject to some limitations. First, it has assumed a simple
split-monopole topology for the magnetic field; while this is not
unrealistic, highly accurate numerical simulations, either in
magnetohydrodynamics or with particle-in-cell approaches, will be
necessary to assess the precise topology of the magnetic field on horizon
scales. Second, the expressions derived here are valid at
$\mathcal{O}(\Omega^6_h)$ and hence may be not very accurate when the BH
spin approaches the extreme value $a_*=1$; this motivates the inclusion
of additional and non-polynomial terms in the perturbative
approach.
We expect that, similar to the Kerr case, these corrections,
self-consistently obtained by enhancing the perturbative approach via
matched asymptotic expansions can significantly improve the accuracy
of the solution close to the extreme regime~\cite{Camilloni2022}. This
will be the subject of future extension of the present work, whose aim
is laying the foundations for a theory-agnostic approach to the BZ
mechanism.
Finally, while mathematically correct, the expressions derived
here have not yet been contrasted with fully nonlinear simulations, which
will actually provide not only the validation of the approach, but also
help in improving the modelling in the near-extremality regime. All of
these aspects will be explored in future works.

\subsection*{Acknowledgements}

We thank Yixuan Ma for his contribution in a preliminary stage of the
project. We are grateful to Christian Ecker for useful advice and with
Laurent Loinard for fruitful discussions.  Support comes from the ERC
Advanced Grant ``JETSET: Launching, propagation and emission of
relativistic jets from binary mergers and across mass scales'' (Grant
No. 884631). F.~C. acknowledges support from the Horizon Europe staff
exchange (SE) programme HORIZON-MSCA2021-SE-01 Grant
No. NewFunFiCO-101086251. L.~R. acknowledges the Walter Greiner
Gesellschaft zur F\"orderung der physikalischen Grundlagenforschung
e.V. through the Carl W. Fueck Laureatus Chair.

\appendix
\section{Grad-Shafranov equation for stationary and axisymmetric
  force-free magnetospheres}
\label{app:BHMAG}

With the Faraday tensor given in Eq.~\eqref{eq:Faraday} and when
considering an axisymmetric and stationary background spacetime, only
the $\phi$-component of Eqs.~\eqref{eq:FFE_2} survives and it is
equivalent to an integrability condition for the field angular
velocity, \ie $\partial_r\Omega_f\, \partial_\theta\Psi =
\partial_\theta\Omega_f\, \partial_r\Psi$, so that that $\Omega_f(r,
\theta) = \Omega_f(\Psi)$. In other words, the angular velocity is
constant along the poloidal magnetic field lines, that are therefore
rigidly rotating.
Similarly, it is possible to show that not every component of the FF
condition in Eq.~\eqref{eq:FFE} is independent. In particular, from the
condition $F_{\mu\nu} J^\mu=0$ it is possible to derive that
$F_{t\mu}J^\mu = -\Omega_f F_{\phi\mu}J^\mu$, which leads to the
integrability condition $\partial_r I \, \partial_\theta\Psi =
\partial_\theta I \, \partial_r\Psi$, that is, $I(r, \theta) =
I(\Psi)$. Stated differently, the toroidal components of the magnetic
field (which are sourced by the poloidal current $I(r, \theta)$), are
constant along the poloidal field lines as measured by an asymptotic
observer.
The last two components of the FF conditions are again not independent,
since $\partial_\theta\Psi \, F_{r\mu} J^\mu = \partial_r \Psi \,
F_{\theta\mu}J^\mu$, so that, together with the integrability conditions,
it is possible to arrive at the \emph{Grad-Shafranov} (GS) equation, that
is, the master equation for stationary and axisymmetric FF
magnetospheres. In the case of a HP-KRZ background, the GS equation takes
a compact form whose structure is similar to the one derived in the Kerr
metric~\cite{Camilloni2022}
\begin{equation}
  \label{eq:GSE}
  \eta_\alpha\partial_\beta\left(\sqrt{-g} \eta^\alpha
  g^{\beta\rho}\partial_\rho\Psi\right)+\frac{\sqrt{-g}
    I(\Psi)}{r^2N^2\sin^2\theta}\dfrac{dI}{d\Psi}=0\, .
\end{equation}
The GS equation is a quasi-linear, second-order, partial differential
equation that must be solved for the flux function $\Psi$, but that also
contains $\Omega_f(\Psi)$ and $I(\Psi)$ as unknown functions of the
primary variable.

Solving the magnetospheric problem requires taking into account the
existence of critical surfaces in the GS equation, where the equation
passes from being second order to be of first order~\cite{Uzdensky2004,
  Uzdensky2005}. Over these critical surfaces, it is necessary to
supplement the GS equation with a set of regularity conditions. In
particular, at the KRZ horizon $r = r_0$, one has the so-called Znajek
condition at the horizon~\cite{Znajek77, Nathanail2014}
\begin{equation}
  \label{eq:ZH}
  I(r = r_0,\theta) =\frac{R_{_{\rm
        M}}\sin\theta}{r\Sigma}\left(\omega-\Omega_f\right)\partial_\theta
  \Psi\,.
\end{equation}
This regularity condition also stems from requiring that the toroidal
magnetic field measured by a normal observer using HP coordinates remains
finite at the BH horizon. On the other hand, in the asymptotic region
$r\to\infty$, it is possible to use the Znajek condition at
infinity~\cite{Uzdensky2004, Nathanail2014}
\begin{equation}
  \label{eq:Zinf}
  I(r,\theta) = \sin\theta~\Omega_f~\partial_\theta \Psi\, , ~~\text{ for }~~
  r\to\infty\, , 
\end{equation}
that can be obtained directly from the GS equation by considering the
limit $r\to \infty$, where it becomes of first order and can be directly
integrated.

In a BH magnetospheric problem, two additional critical surfaces arise,
called the inner and outer \emph{light surface} (ILS/OLS),
respectively. While the OLS is the direct analogue of the light cylinder
in pulsar magnetospheres, the presence of the ILS is a purely
general-relativistic effect that is related to the frame-dragging and,
indeed, the ILS is always confined within the BH
ergosphere~\cite{Komissarov2005}. The locations of the ILS and the OLS
are given by the roots of the condition $\eta^\mu
\eta_\mu=0$~\cite{Gralla2014}, corresponding to the points where the
frames corotating with the magnetic-field lines reach the speed of
light. When this condition is valid, the GS equation becomes a
first-order PDE and, for the solutions to be continuous at the light
surfaces, it is necessary to impose a Robin-type condition also referred
to as the ``reduced stream equation''~\cite{Uzdensky2005, Armas2020,
  Camilloni2022}
\begin{equation}
  \label{eq:GS_reduced}
  \left(\eta_\mu\partial_\nu\eta^\mu\right) g^{\nu\rho}\partial_\rho\Psi
  + \frac{I(\Psi)}{r^2N^2\sin^2\theta}\dfrac{dI}{d\Psi}=0\, .
\end{equation}
Even though the reduced stream equation will not play a role in the
present discussion, we still report it for completeness, as it plays a
fundamental role in the perturbative construction of higher-orders
magnetospheric models via the matched asymptotic-expansion scheme (see,
\eg \cite{Armas2020, Camilloni2022}).

\section{Blandford--Znajek mechanism}
\label{app:BZ}

The expressions for the rates of extraction of energy and angular
momentum in a generic KRZ spacetime can be derived -- in analogy with
what done in GR -- by considering the flux of the electromagnetic
energy-momentum tensor associated, \ie
\begin{align}
  & {T^r}_t = \Omega_f(\Psi)\, I(\Psi)\, \partial_\theta
  \Psi(r, \theta)/\sqrt{-g}\, , \\
  & {T^r}_\phi = -I(\Psi) \, \partial_\theta \, \Psi(r, \theta)/\sqrt{-g}\, .
\end{align}
Upon integrating over a sphere of constant coordinate radius, one can
define~\cite{Gralla2014}
\begin{align}
  \label{eq:BZ}
  & \mathcal{\dot{E}}_{_{\rm BZ}}(r) := \left(\frac{d \mathcal{E}}{dt}\right)_{_{\rm BZ}}  
  = - \int \Omega_f(\Psi)\, I(\Psi)\, d\Psi\, , \\
  & \mathcal{\dot{L}}_{_{\rm BZ}}(r) := \left(\frac{d \mathcal{L}}{dt}\right)_{_{\rm BZ}}  
  = - \int I(\Psi)\, d\Psi\, , 
\end{align}
as the rates of change of the energy and the angular momentum
contained within the spherical surface of radius $r$, flowing outward
or inward along the poloidal field lines given by the constant $\Psi$
contours.
From the expressions above one immediately recognises that an
extraction of energy is associated to an extraction of angular
momentum, with the proportionality factor being given by the angular
velocity of field lines $\delta \mathcal{E}_{_{\rm BZ}} = \Omega_f \delta
\mathcal{L}_{_{\rm BZ}}$~\cite{Blandford1977}.

In practice, it is convenient to perform the integral at the event
horizon, \ie for $r = r_0$, where the Znajek condition~\eqref{eq:ZH},
can be used to express Eq.~\eqref{eq:BZ} as
\begin{equation}
  \label{eq:BZ1}
  \mathcal{\dot{E}}_{_{\rm BZ}}=-\int\frac{R_{_{\rm
        M}}\sin\theta}{r\Sigma}
  \Omega_f(\Psi)\left[\Omega_h-\Omega_f(\Psi)\right]
  d\Psi\bigg\vert_{r_0}\, ,
\end{equation}
where $d\Psi := \partial_r \Psi\, dr + \partial_\theta \Psi\,
d\theta$.

In order to relate the energy computed at the horizon via the
integral~\eqref{eq:BZ1}, with that reaching spatial infinity, we
introduce the BZ power defined as $P_{_{\rm BZ}} := -
\mathcal{\dot{E}}_{_{\rm BZ}}$, so that energy is extracted from the
BH via the BZ mechanism ($P_{_{\rm BZ}} > 0, \,
\mathcal{\dot{E}}_{_{\rm BZ}} < 0$) if and only if $\Omega_h >
\Omega_f > 0$, namely, when the magnetic-field lines rotate slower
than the BH horizon. From a mechanical point of view, this result can
be interpreted as if the field lines oppose an inertial resistance to
the frame-dragging induced by the BH, in analogy with the mechanical
Penrose process~\cite{Lasota2014,Kinoshita2017}. The maximum
energy-extraction rate is reached for $\Omega_f = \Omega_h/2$ and, as
shown below, is obtained for a monopolar magnetospheric topology. It
also follows that $\mathcal{\dot{E}}_{_{\rm BZ}} \leq \Omega_h
\mathcal{\dot{L}}_{_{\rm BZ}}$, in agreement with the second law of BH
mechanics~\cite{Blandford1977}.

We should remark that the approach we have followed so far is valid in a
generic axisymmetric and stationary spacetime and that the corresponding
explicit expressions obtained are expressed for a HP-KRZ formulation,
which itself is rather generic. Hereafter, we will concentrate our
attention on what arguably is the most interesting quantity from an
astrophysical point of view, namely, the extracted power $P_{_{\rm
    BZ}}$.
This quantity is clearly a function of the BH spin and its
dependence from $\Omega_h$ has been computed analytically by means of
perturbative approaches~\cite{Tanabe2008, Armas2020, Camilloni2022} or
via numerical simulations~\cite{Tchekhovskoy2010, Meringolo2025a}. The
general functional form of this power in GR can be expressed as
\begin{equation}
  \label{eq:BZ2}
  P_{_{\rm BZ}}(\Omega_h) = \kappa\, (2\pi \Psi_h)^2 \, \Omega_h^2 \, f(\Omega_h)\,.
\end{equation}
where $\kappa$ is a geometrical factor associated to the topology of
the magnetic field around the BH, $(2\pi\Psi_h)$ is the magnetic flux
at the horizon, and $f(\Omega_h)$ is the so-called \emph{high-spin
factor} and essentially encodes the spin dependence of the BZ
power. What is known in GR is that $f(\Omega_h)\approx 1$ for
slow-spinning BHs, $M \Omega_h \ll 1$, so that one recovers the
quadratic scaling of the BZ power originally derived by Blandford and
Znajek, $ P_{_{\rm BZ}} \propto \Omega_h^2$~\cite{Blandford1977}. On
the other hand, for high-spinning BHs, when $M \Omega_h \simeq 1$,
$f(\Omega_h)$ contains terms in a power expansion in $\Omega_h$,
together with logarithmic contributions~\cite{ Camilloni2022}. In
particular, restricting to an expression at order
$\mathcal{O}(\Omega^6_h)$
\begin{equation}
  \label{eq:f_GR}
    f(\Omega_h) = 1 + {\alpha} M^2\Omega^2_h + {\beta} M^4\Omega^4_h +
    \mathcal{O}(\Omega_h^5)\, ,
\end{equation}
where the coefficients in the expansion are given by ${\alpha} =
8{(67 - 6\pi^2)} / {45}\simeq 1.38$~\citep{Tanabe2008} and ${\beta}
\simeq -11.25$~\citep{Camilloni2022}.

Our main result, Eq.~\eqref{eq:BZKRZ}, makes clear that the
\emph{ansatz}~\eqref{eq:BZ2} is true not only in GR but also in
generic KRZ spacetimes, where, of course, the functional form of the
high-spin factor will need to be extended as a function of the KRZ
parameters $f_{_{\rm KRZ}} = f_{_{\rm KRZ}}(\Omega_h, \varrho,
a_1,\dots, b_1,\dots)$. At the same time, in Sec.~\ref{sec:BZJET} we
showed how to exploit the richer functional dependence to learn about
strong-gravity near the BH from the measurements of the radiated BZ
power.

\section{Perturbative approach to the magnetospheric problem}
\label{sec:ffepert}

In order to derive an explicit expression for the BZ luminosity one has
to solve the GS equation and its regularity conditions so as to obtain
information about the field variables $(\Psi, \Omega_f, I)$ in the HP-KRZ
background. To this scope, we follow the analytic approach originally
devised by Blandford and Znajek~\cite{Blandford1977}, that consists in
solving the magnetospheric problem perturbatively in the low-spin regime
$a_* \ll 1$ for KRZ BHs. Such an approach is built on a physical
intuition: it is reasonable to expect that in the low-spin regime the
magnetic field around the BH will be similar to that of a static
electrovacuum solution, hence $F_{\mu\nu} \sim \mathcal{O}(a_*^0) =
\mathcal{O}(1)$. At the same time, it is well-known that in stationary
backgrounds the dragging of inertial frames induces electric fields and
the associated potential gaps~\cite{Wald:74bh}. These gaps can be natural
sites of particle production via pair-cascade, ultimately leading to a
replenishment of plasma in the BH environment~\cite{Blandford1977}. A
plasma-filled magnetosphere is thus established around the BH and can be
described by a current-density that scales as $J^{\mu} \sim
\mathcal{O}(a_*)$, so that the FF constraint is given by $F_{\mu\nu}
J^\mu \sim \mathcal{O}(a_*)$, that is, it is linear in the BH spin.

One explicit way of realising this condition is by assuming the following
scaling for the magnetospheric field variables: $\Psi \sim
\mathcal{O}(a_*^0)$, $\Omega_f \sim \mathcal{O}(a_*)$, and $I \sim
\mathcal{O}(a_*)$. Within this perturbative framework it is then possible
to start from any magnetic flux that is an electrovacuum solution in the
static limit and assume that the magnetic-field lines acquire an angular
velocity and additional toroidal components proportional to the BH spin
via frame-dragging.

Assuming that the variables are expanded as
\begin{align}
  &\Psi = \psi_0 + \sum_{n=1}^{\infty} a_*^n \,\psi_n\,, \\
  &\Omega_f = \sum_{n=1}^{\infty} a_*^n \,\omega_n \,, \\
  &I = \sum_{n=1}^{\infty} a_*^n \, i_n\,,
\end{align}
with $n\geq1$, the FFE equations can be solved order-by-order in the $a_*
\ll 1$ expansion. In particular, the GS equation at the $n$-th order
reads as
\begin{equation}
  \label{eq:GS_L}
  \mathscr{L} \psi_n (r, \theta) = \mathcal{S}_n\left(r, \theta;
  \psi_{n-1}, \omega_{n-1}, i_{n-1}\right)\,,
\end{equation}
where $\mathscr{L}$ is a partial-differential differential operator whose
expression in the HP-KRZ background is given by
\begin{equation}
  \label{eq:op_L}
  \mathscr{L} : = \frac{1}{R_{_{\rm
        B}}\sin\theta}\partial_r\left[\frac{1}{R_{_{\rm
          B}}}\left(1-\frac{\tilde R_{_{\rm
          M}}}{r}\right)\partial_r\right] +
  \frac{1}{r^2}\partial_\theta\left(\frac{1}{\sin\theta}
  \partial_\theta\right)\, .
\end{equation}
The functions $R_{_{\rm M}}$ and $R_{_{\rm B}}$ appearing above have
explicit expression presented in
Eqs.~\eqref{eq:RB_expan}--\eqref{eq:RM_expan}, where we have defined
$\tilde{R}_{_{\rm M}} := R_{_{\rm M}}(a_*=0)$. Furthermore, the source
term $\mathcal{S}_n$ depends on the previous perturbative corrections and
on the background metric.

In order to maintain the treatment at an analytic level, we make the
assumption that the solution and the source term are separable in the
radial and polar coordinates, \ie
\begin{align}
  \label{eq:sep_ansatz}
  & \psi_n(r, \theta) = \mathcal{R}_{n\ell}(r) \Theta_{\ell}(\theta) \,, \\
  & \mathcal{S}_n(r, \theta) = s_{n\ell}(r)\Theta_\ell(\theta)/(R_{_{\rm B}}\sin\theta)\,,
\end{align}
so that the differential operator~\eqref{eq:op_L} in the HP-KRZ
background can be written as
\begin{equation}
  \label{eq:op_L_split}
  \mathscr{L} =
  \frac{1}{R_{_{\rm B}} \sin \theta}\mathscr{L}_r^{_{(\ell)}} +
  \frac{1}{r^2} \mathscr{L}_\theta^{_{\rm (\ell)}} \,,
\end{equation}
with
\begin{align}
  \label{eq:Lr}
  &\mathscr{L}_r^{_{(\ell)}}:=\partial_r\left[\frac{1}{R_{_{\rm
          B}}}\left(1-\frac{\tilde R_{_{\rm
          M}}}{r}\right)\partial_r\right]-\frac{\ell(\ell + 1) R_{_{\rm
        B}}}{r^2}\,,\\
  \label{eq:Ltheta}
  &\mathscr{L}_\theta^{_{(\ell)}}:=\partial_\theta
  \left(\frac{1}{\sin\theta}\partial_\theta \right) +
  \frac{\ell(\ell + 1)}{\sin\theta} \,,
\end{align}
and where $\ell$ is an integer that is inherited from the relation
between the differential operator~\eqref{eq:op_L_split} with the Legendre
equation. In this way, the GS operator~\eqref{eq:op_L} takes the form
\begin{equation}
  \mathscr{L}\psi_n(r, \theta) = \frac{\Theta_\ell(\theta)}{R_{_{\rm
        B}}\sin\theta} \left[ \mathscr{L}_r^{_{(\ell)}} \mathcal{R}_{n
      \ell}(r) \right]+\frac{ \mathcal{R}_{n \ell}(r)}{r^2} \left[
    \mathscr{L}_\theta^{_{(\ell)}}\Theta_\ell(\theta) \right]\, ,
\end{equation}
and, provided $\Theta_\ell(\theta)$ is an angular eigenfunction
satisfying $\mathscr{L}_\theta^{_{(\ell)}}\Theta_\ell(\theta)=0$, the GS
equation~\eqref{eq:GS_L} reduces to a set of ODEs for the radial parts
alone~\cite{Camilloni2022}
\begin{equation}
    \mathscr{L}_{r}^{_{(\ell)}}\mathcal{R}_{n \ell}(r)=s_{n\ell}(r)\, .
\end{equation}
Furthermore, in the case of a zero source term, the radial and
angular eigenfunctions of the operators~\eqref{eq:Lr} and
\eqref{eq:Ltheta}, \ie $\varphi_{\ell}(r)$ and $\Theta_{\ell}(\theta)$,
are the solutions of the Maxwell equations in a static HP-KRZ background.

An explicit solution of the radial-operator equation
$\mathscr{L}_r^{_{(\ell)}} \varphi_\ell(r) = 0$ is difficult to obtain
analytically in the general case, as the radial operator depends on all
KRZ parameters via $R_{_{\rm M}}$ and $R_{_{\rm B}}$. Closed-form
expressions for $\varphi_\ell$ are known in specific settings such as the
Kerr case, where the equation reduces to a Papperitz-Riemann
equation~\cite{Gralla2016, Camilloni2022}, or for metrics sharing the
same line-element structure of the Kerr-Newman spacetime, where the
equation becomes a Heun equation~\cite{Camilloni2023}. In practice, in
the following we will deal with the solution of the radial part of the
magnetic flux via a direct numerical solution. On the other hand, the
angular eigenfunctions obey the same equation as in the Schwarzschild
spacetime, $\mathscr{L}_\theta^{_{(\ell)}} \Theta_\ell(\theta) = 0$, and
their expression can be easily given in closed form in terms of the
Gegenbauer polynomials~\cite{Petterson:1975, Gralla2016}. For later
convenience, we here report only the first functions
\begin{align}
  \label{eq:Theta}
    &\Theta_0:=1-\cos\theta\, , \\
    &\Theta_1:=\sin^2\theta = \Theta_3/(1-5\cos^2\theta)\, , \\
    &\Theta_2:=\cos\theta\sin^2\theta = \Theta_4/\left[1-(7/3)\cos^2\theta\right]\, .
\end{align}
Obviously, the angular eigenfunctions form a complete orthogonal basis
over which to expand the polar dependence of the various quantities.

\subsection{Quasi-universality of the split-monopole magnetosphere}

As mentioned above, the starting point of the perturbative approach
consists in finding an electrovacuum solution in the limit $a_*=0$, so
that, adopting the aforementioned scaling one has
\begin{align}
  \Psi(r, \theta) &= \psi_0(r, \theta)+\mathcal{O}(a_*^2)\,,\\
  I(\Psi) &= \Omega_f(\Psi)=\mathcal{O}(a_*)\,.
\end{align}
The GS equation is then given by the static electrovacuum equation
\begin{equation}
  \label{eq:Lpsi0}
  \mathscr{L}\psi_0 = 0 + \mathcal{O}(a_*^2)\,,
\end{equation}
that is solved by any of the eigenfunctions described in the previous
subsection. 

The simplest of the eigenfunctions of Eq.~\eqref{eq:Lpsi0} is
associated to a split-monopole configuration, obtained by demanding
$\Psi(r,\theta)$ to remain finite for $r\to\infty$ at fixed polar
angles $\theta$~\cite{Grignani2019}.
The split-monopole field does not depend on the radius and its
explicit expression is given by
\begin{equation}
  \label{eq:psi0}
  \psi_0 := \Psi_h\,{\rm sign}(\cos\theta)\Theta_0(\theta)\, , 
\end{equation}
with $\Psi_h$ being the magnetic flux function at the horizon, that in
this perturbative construction is held constant
(see~\cite{Chatterjee2023} for an alternative treatment where the
magnetic flux at the horizon is computed from the accretion process in a
magnetically-arrested disk state). We should remark that even if the
split-monopole field appears as an idealised set-up, recent simulations
have shown many non-equilibrium magnetic field around a BH (for instance
inherited by the progenitors in neutron-star mergers) ultimately evolving
into a split-monopole topology~\cite{Bransgrove2021, Selvi2025}, renewing
the interest in this configuration. Furthermore, the reversal of the
magnetic-field polarity across the equatorial plane, that is encoded in
the prefactor ${\rm sign}(\cos\theta)$, implies that the split-monopole
is sustained by an equatorial current-sheet\footnote{Hereafter, we will
limit our discussion to the northern BH hemisphere, \ie to $\theta\in[0,
  \pi/2)$. The inclusion of the southern hemisphere is trivial and
  amounts to a sign change in Eq.~\eqref{eq:psi0} (reflection
  symmetry)~\cite{Gralla2014, Armas2020}.}. This is a natural location
for magnetic reconnection to occur, with particles being accelerated to
large Lorentz factors and plasmoid being formed in a highly dynamical
plasmoid chain~\cite{Nathanail2020, Bransgrove2021, Nathanail2021b,
  Camilloni2025, Meringolo2025a}. Finally, as we will demonstrate below,
magnetospheres with a split-monopole topology always exist for any BH
spacetime within the HP-KRZ class.

Beyond the leading order, the split-monopole fields receives additional
corrections induced by the frame dragging. The detailed computation to
solve order-by-order the GS equation and its regularity conditions in the
HP-KRZ metric is not significantly different from the computation
performed in the Kerr scenario, and we refer the reader to the previous
literature for details~\cite{Tanabe2008, Armas2020, Camilloni2022,
  Camilloni2023}, while we here limit ourselves to reporting the
structure of the solutions. More specifically, the magnetic flux function
can be expanded around a split-monopole as
\begin{equation}
\Psi = \Psi_h(1-\cos\theta+a_*^2\psi_2+a_*^4\psi_4)+\mathcal{O}(a_*^5)\,, 
\end{equation}
so that, when exploiting the ansatz~\eqref{eq:sep_ansatz}, the
perturbative expression reads~\cite{Camilloni2022}

\begin{widetext}
\begin{equation}
  \label{eq:Psi}
  \begin{split}
    \Psi=\Psi_h&\left[1-\cos\theta +
      a_*^2~\mathcal{R}_{22}(r)\Theta_2(\theta) +
      a_*^4~\Big(\mathcal{R}_{42}(r)\Theta_2(\theta) +
      \mathcal{R}_{44}(r)\Theta_4(\theta)\Big)\right] +
    \mathcal{O}(a_*^5)\, ,
  \end{split}
\end{equation}
\end{widetext}

\noindent where the angular harmonics are given explicitly in
Eq.~\eqref{eq:Theta}, and the radial coefficients $\mathcal{R}_{22}(r)$,
$\mathcal{R}_{42}(r)$ and $\mathcal{R}_{42}(r)$ obey a set of ordinary
differential equations of the type $\mathscr{L}_r^{_{(\ell)}}
\mathcal{R}_{n, \ell}(r) = s_{n, \ell}(r)$, that we report explicitly
together with the expressions of the source terms for the KRZ background
in Appendix~\ref{app:shooting}.

The expression for the poloidal current $I(\Psi)$, related to toroidal
magnetic fields, and the angular velocity of the field lines
$\Omega_f(\Psi)$, associated to gravity-induced electric fields, are
determined by combining the Znajek conditions at the horizon and at
infinity, \ie Eqs.~\eqref{eq:ZH} and~\eqref{eq:Zinf}~\cite{Armas2020,
  Camilloni2022}. The corresponding expansions are given by

\begin{widetext}
\begin{align}
  \label{eq:I_Psi}
  I(\Psi) &=\Psi_h\frac{a_*}{8}\frac{(1 + \varrho)^2}{M}\Theta_1(\theta)~\left[1
    + \frac{a_*^2}{4}\left(1-2\varrho + \varrho^2 +
    8\mathcal{R}_{22}(r)-2\left(4\mathcal{R}_{22}(r) +
    \mathcal{R}_{22}^{h}-\frac{(1 +
      \varrho)^2}{4}\right)\Theta_1(\theta)\right)\right] +
  \mathcal{O}(a_*^4)\, , \nonumber \\
    &=\frac{1}{2}\Psi_h\Omega_h\Theta_1(\theta)\,\left[1
    + 32\frac{M^2\Omega_h^2}{(1+\varrho)^4}\left(
    \mathcal{R}_{22}(r)-\frac{1}{4}\left(4\mathcal{R}_{22}(r) +
    \mathcal{R}_{22}^{h}-\frac{(1 +
      \varrho)^2}{4}\right)\Theta_1(\theta)\right)\right] +
  \mathcal{O}(\Omega_h^4)\, , 
\end{align}

\noindent and

\begin{align}
\label{eq:Omega_Psi}
  \Omega_f (\Psi) &=\frac{a_*}{8}\frac{(1 + \varrho)^2}{M}\left[1 -
    \frac{a_*^2}{4}\left(1 - 2\varrho - \varrho^2 -
    2\left(\mathcal{R}_{22}^{h} - \frac{(1 +
      \varrho)^2}{4}\right)\Theta_1(\theta)\right)\right] +
  \mathcal{O}(a_*^4) \nonumber \\
  &=\frac{1}{2}\Omega_h\left[1 -
    8\frac{M^2\Omega_h^2}{(1+\varrho)^4}\left(\mathcal{R}_{22}^{h} - \frac{(1 +
      \varrho)^2}{4}\right)\Theta_1(\theta)\right] +
  \mathcal{O}(\Omega_h^4)\, .
\end{align}
\end{widetext}

\noindent with $\mathcal{R}_{n, \ell}^{h}$ marking the value of the
radial magnetic fluxes at the horizon, and where the second expressions
for the right-hand side in Eqs.~\eqref{eq:I_Psi} and \eqref{eq:Omega_Psi}
are given in terms of $\Omega_h$ rather than $a_*$. 

Note that the leading-order term in the expansion for the angular
velocity of the field lines in~\eqref{eq:Omega_Psi} is consistent with
the rotation law expected for a monopolar field, $\Omega_f = \Omega_h/2 +
\mathcal{O}(\Omega_h^3)$~\cite{Blandford1977}, when the spin parameter is
converted into the BH angular velocity
\footnote{It is useful to stress that the condition $\Omega_f\propto
\Omega_h$ follows from the scaling underlying the perturbative
approach employed here.
Alternative approaches do not necessarily imply a relation between
these quantities \cite{Menon2005, Brennan2013, Camilloni2020a,
  Camilloni2020b, Menon2024}, but none of the corresponding FFE
solutions is associated to an active BZ process and to BH energy
extraction.
By contrast, FF configurations obtained perturbatively but with
non-monopolar boundary conditions (\eg parabolic~\cite{Blandford1977},
hyperbolic~\cite{Gralla2016} or vertical~\cite{East2018}) are
typically characterised by more complicated proportionality factors
between $\Omega_f$ and $\Omega_h$ that do not maximise energy
extraction.}.
Perturbative corrections to $I(\Psi)$ and $\Omega_f (\Psi)$ of order
$\mathcal{O}(a_*^4)$ or higher can be derived by complementing the
perturbation theory via matching asymptotic expansions at the
OLS~\cite{Armas2020, Camilloni2022}. Even though these extensions to
higher orders are very interesting, they are not needed in the present
estimate of the BZ luminosity in KRZ spacetimes at
$\mathcal{O}(a_*^6)$, but they will be explored in future work.

\section{Shooting method for split-monopole magnetosphere}
\label{app:shooting}

%
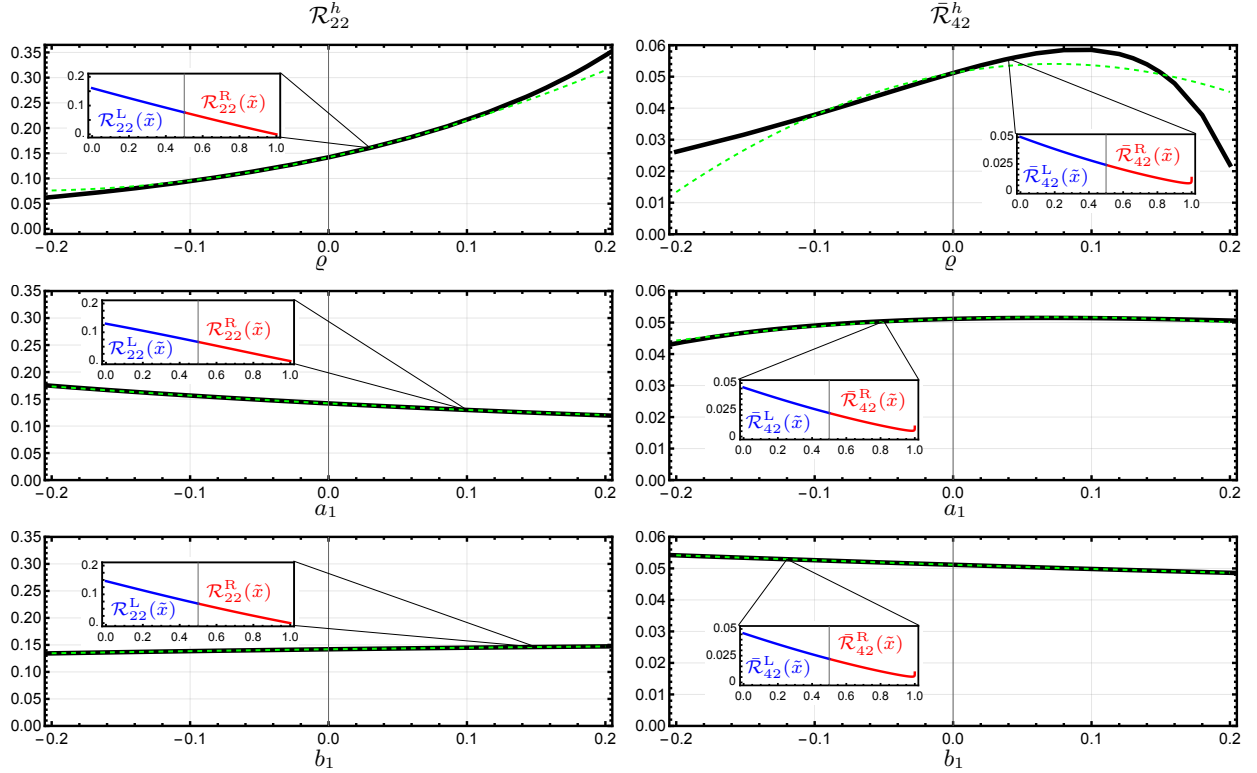
\begin{figure*}[t!]
  \centering
  \includegraphics[width=0.45\textwidth]{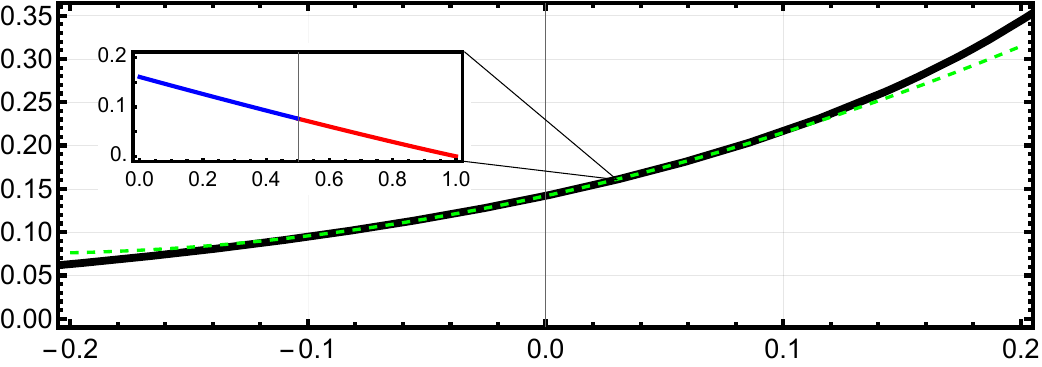}~~~\includegraphics[width=0.45\textwidth]{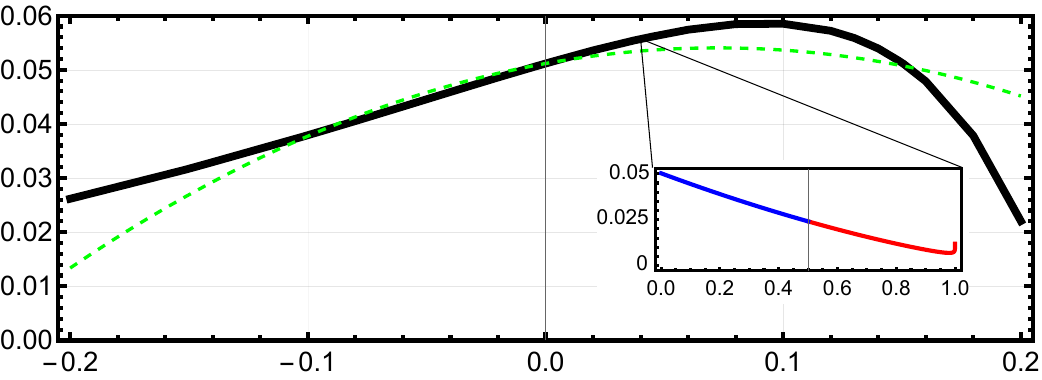}\\\vspace{3mm}
  \includegraphics[width=0.45\textwidth]{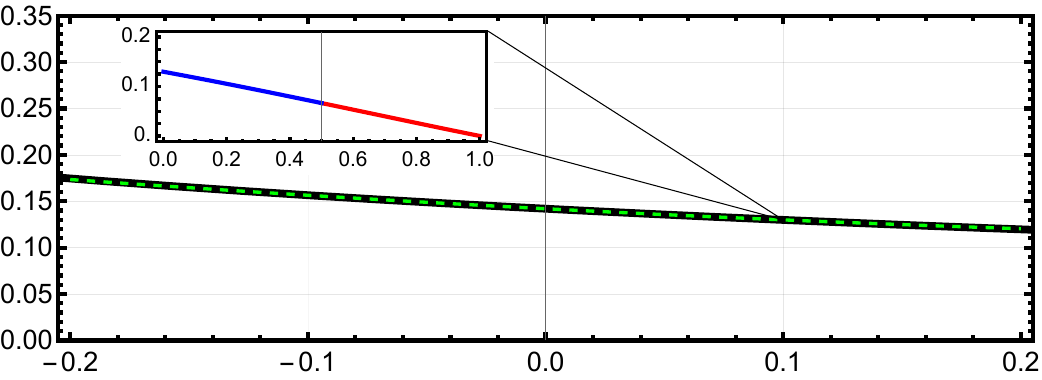}~~~\includegraphics[width=0.45\textwidth]{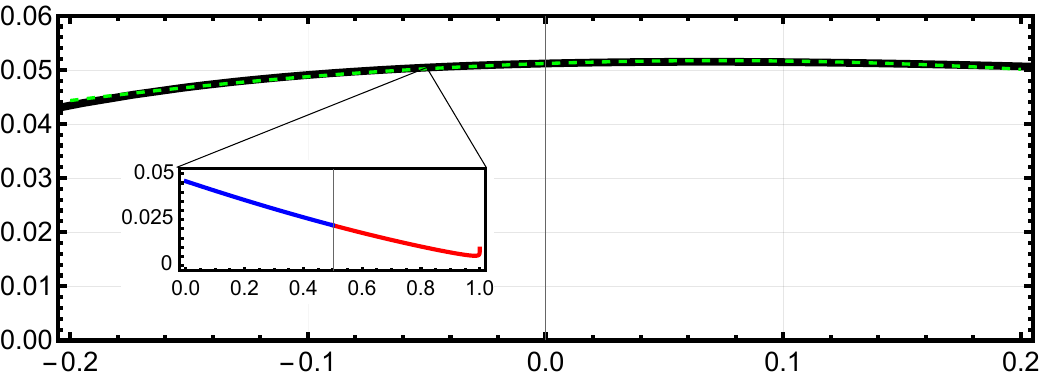}\\\vspace{3mm}
  \includegraphics[width=0.45\textwidth]{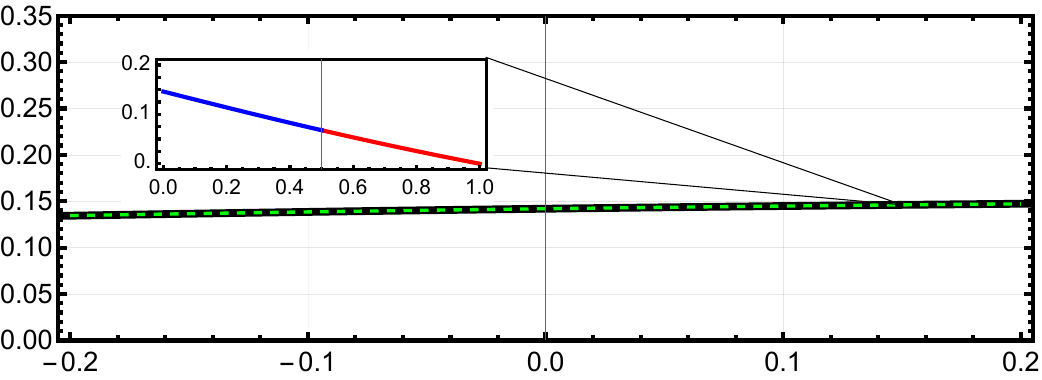}~~~\includegraphics[width=0.45\textwidth]{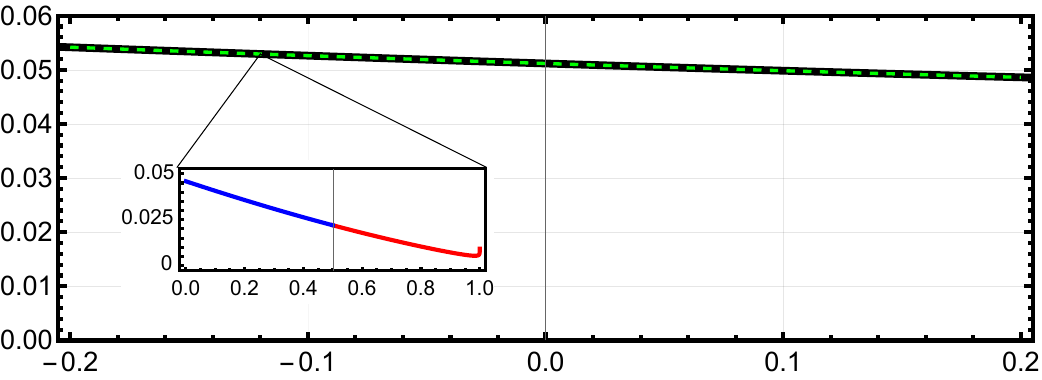}\\
   \begin{picture}(0, 0)
     \put(-120, 285){${\mathcal{R}}_{22}^{h}$}
     \put(-118, 193){$\varrho$}
     \put(-118, 99){$a_1$}
     \put(-118, 4){$b_1$}
     \put(-200, 246){\scriptsize \color{blue}{${\mathcal{R}}_{22}^{\rm L}(\tilde x)$}}
     \put(-161, 253){\scriptsize \color{red}{${\mathcal{R}}_{22}^{\rm R}(\tilde x)$}}
     \put(-196, 160){\scriptsize \color{blue}{${\mathcal{R}}_{22}^{\rm L}(\tilde x)$}}
     \put(-159, 167){\scriptsize \color{red}{${\mathcal{R}}_{22}^{\rm R}(\tilde x)$}}
     \put(-195, 61){\scriptsize \color{blue}{${\mathcal{R}}_{22}^{\rm L}(\tilde x)$}}
     \put(-159, 68){\scriptsize \color{red}{${\mathcal{R}}_{22}^{\rm R}(\tilde x)$}}
     \put(115, 285){$\bar{\mathcal{R}}_{42}^{h}$}
     \put(120, 193){$\varrho$}
     \put(120, 99){$a_1$}
     \put(120, 4){$b_1$}
     \put(185, 233){\scriptsize \color{red}{$\bar{\mathcal{R}}_{42}^{\rm R}(\tilde x)$}}
     \put(150, 225){\scriptsize \color{blue}{$\bar{\mathcal{R}}_{42}^{\rm L}(\tilde x)$}}
     \put(81, 140){\scriptsize \color{red}{$\bar{\mathcal{R}}_{42}^{\rm R}(\tilde x)$}}
     \put(45, 132){\scriptsize \color{blue}{$\bar{\mathcal{R}}_{42}^{\rm L}(\tilde x)$}}
     \put(80, 48){\scriptsize \color{red}{$\bar{\mathcal{R}}_{42}^{\rm R}(\tilde x)$}}
     \put(45, 39){\scriptsize \color{blue}{$\bar{\mathcal{R}}_{42}^{\rm L}(\tilde x)$}}
   \end{picture}
  \caption{Dependence of the horizon values ${\mathcal{R}}_{22}^{h}$
    and $\bar{\mathcal{R}}_{42}^{h}$ [see Eq.~\eqref{eq:barR} for a
      definition] on the KRZ parameters $\varrho$ (top panel) $a_1$
    (middle panel), and $b_1$ (bottom panel); the corresponding values
    for the Kerr spacetime correspond can be found when the parameters
    are equal to zero. The various insets illustrate an example of the
    solutions obtained via the shooting method in the compact
    coordinate $\tilde{x}\in[0, 1]$ and where the blue line represents
    the solution from the horizon $\tilde{x}=0$ up to the matching
    point $\mathcal{R}^{\rm L}$, while the red line the solution from
    the asymptotic infinity $\tilde{x}=1$ to the matching point
    $\mathcal{R}^{\rm R}$. Finally, shown with green dashed lines are
    the analytic approximations in of the various quantities and
    highlighting the very good match [see
      Eqs.~\eqref{eq:R22R42appr}--\eqref{eq:R22R42appr_last}].}
\label{fig1}
\end{figure*}

Here we describe how the equations obtained perturbatively can be
easily integrated via shooting. Besides supporting our analytic
computation in Sec.~\ref{sec:BZ_plots}, we envisage this Appendix
presents an interesting and easy-to-apply method to derive
BH magnetospheric solutions in GR and beyond.

In the perturbative approach, detailed in Sec.~\ref{sec:ffepert}, the
magnetospheric field variables are expanded according to
Eq.s~\eqref{eq:Psi} and~\eqref{eq:I_Psi}, and the GS equation
reduces to a set of ODEs of the general form
\begin{equation}
    \mathscr{L}_{r}^{_{(\ell)}}\mathcal{R}_{n \ell}(r)=s_{n\ell}(r)\, .
\end{equation}
Here $\mathcal{R}_{n, \ell}$ is the radial part of the magnetic flux at
the $n$th order in the perturbation theory and the differential
operator is explicitly given in Eq.~\eqref{eq:Lr} for the KRZ
background.
As discussed in Sec.~\ref{sec:ffepert} we stress that the character of
the differential operator $\mathscr{L}_{r}^{_{(\ell)}}$ is only known
once a specific set of the KRZ parameters is chosen.

More specifically, at the second order in the low-spin expansion, $n=2$,
the GS equation becomes $\mathscr{L}_r^{_{(2)}} \mathcal{R}_{22}(r) =
s_{22}(r)$, the explicit expressions for the source term being
\begin{equation}
  s_{22}(r)=-\frac{2 \tilde{R}_{_B} \tilde{R}_{_M} \left[ r^2-{4
      M^2}/{(\varrho +1)^2}\right ]}{r^4 (r-\tilde{R}_{_M})}\, .
\end{equation}
At the fourth order in the perturbation theory, $n=4$, the GS equation
can be split into two independent equations of the kind
$\mathscr{L}_r^{_{(2)}} \mathcal{R}_{42}(r) = s_{42}(r)$ and
$\mathscr{L}_r^{_{(2)}} \mathcal{R}_{44}(r) = s_{44}(r)$.

Given that the quantity $\mathcal{R}_{44}(r)$ does not enter in
Eq.~\eqref{eq:BZKRZ}, we only focus on the first equation, whose source
term is
\begin{widetext}
\begin{equation}
\begin{split}
  &s_{42}(r)=-\frac{3}{14}\frac{ \tilde{R}^2_{_{\rm M}} \Big[64 M^4+r^4
      (3 \mathcal{R}_{22}^{h}-5) (\varrho +1)^4\Big]}{r^6 (\varrho +1)^4
    \left(r-\tilde{R}_{_{\rm M}}\right)^2}
  +\frac{4}{7}\frac{\tilde{R}_{_{\rm M}} \left(4 M^2-r^2 (\varrho
    +1)^2\right)+21 r \left(r \tilde{\tilde{{R}}}_{_{\rm M}}-2
    M^2\right)}{r^4 (\varrho +1)^2 \left(r-\tilde{R}_{_{\rm
        M}}\right)}\mathcal{R}_{22}(r) \\
  &+\frac{3\tilde{R}_{_{\rm
        M}}[16M^2(8 M^2 \tilde{R}_{_{\rm M}}- 12 M^2 r+ r^3 (\varrho
      +1)^2)-3 r^5 (\varrho +1)^4]}{112 M^2 r^4 (\varrho +1)^2
    \tilde{R}_{_{\rm B}}^2 \left(r-\tilde{R}_{_{\rm M}}\right)}
  \mathcal{R}_{22}'(r)
  +\frac{3\left(r^2 (\varrho +1)^2-8 M^2 \right)^2 \tilde{R}_{_{\rm
        M}}'+6 r^4 (\varrho +1)^4}{112 M^2 r^2 (\varrho +1)^2
    \tilde{R}_{_{\rm B}}^2 \left(r-\tilde{R}_{_{\rm M}}\right)}
  \mathcal{R}_{22}'(r) \\
  &+\frac{3}{14}\frac{r^5 (\varrho +1)^4+ 128
    M^4\tilde{R}_{_{\rm M}}}{r^5 (\varrho +1)^4 \left(r-\tilde{R}_{_{\rm
        M}}\right)^2}
    -\frac{3\tilde{R}_{_{\rm M}}\Big[128 M^4 r^2 (\varrho +1)^2-4 M^2 r^2
        (3 \mathcal{R}_{22}^{h}-1) (\varrho +1)^+r^4 (\varrho
        +1)^6\Big]}{56 M^2 r^3 (\varrho +1)^4 \left(r-\tilde{R}_{_{\rm
          M}}\right)^2}\, .
  \end{split}
\end{equation}
\end{widetext}

In the expression above the horizon value $\mathcal{R}_{22}^{h}$ enters
the source term via $I(\Psi)$ and $\Omega_f(\Psi)$, as defined in
Eq.~\eqref{eq:I_Psi}, whereas ${R}_{_{\rm M}}(r) = \tilde{{R}}_{_{\rm
    M}}+a_*^2\tilde{\tilde{{R}}}_{_{\rm M}}+\mathcal{O}(a_*^4)$.

In order to solve the equations explicitly, it is convenient to
compactify the integration domain by adopting the radial coordinate
$\tilde{x}:=1-r_0/r$, such that $\tilde x\in[0, 1]$, with the horizon
and the infinity respectively mapped to $\tilde x=0$ and the
infinity to $\tilde x=1$. Notice that $\tilde x$ is also a natural
coordinate typically adopted in the KRZ framework~\cite{Konoplya2021}.

So far, with the (rather futile) aim of making the expressions as compact
as possible, we restrict to the three coefficients considered in this
work, by setting $\varrho, a_1, b_1\neq0$. Notwithstanding this, we
stress that all the expressions that follow can be easily generalised to
account for other and more general combinations of the KRZ parameters.

From the equations above, the near-horizon behaviour $\tilde x\to0$
can be inferred via Frobenius expansion. The first terms of the
expansions explicitly read
\begin{widetext}
\begin{equation}
  \label{eq: exp1}
  \begin{split}
     \mathcal{R}_{22}&=\mathcal{R}_{22}^{h}+\frac{(b_1+1)^2
       \left[6 \mathcal{R}_{22}^{h} (a_1-2 \varrho +1)-(\varrho
       +1)^2\right]}{(a_1-2 \varrho +1)^2} \tilde{x}+\mathcal{O}(\tilde x)
     \\
     \mathcal{R}_{42}=&\mathcal{R}_{42}^{h}+\frac{\left(b_1+1\right)^2}{112 \left(a_1-2
   \varrho +1\right)^4} \Bigg[2 \left(a_1-2 \varrho +1\right) \Big(-84 \left(b_1+1\right)^2 \left(6
       \mathcal{R}_{22}^{h} \left(a_1-2 \varrho +1\right)-(\varrho +1)^2\right)
       \\
       \phantom{=}&+(\varrho +1)^2 \big(-2 a_1 \left(6
   \mathcal{R}_{22}^{h} (7-4 \varrho )+(2 \varrho +3) \varrho ^2+97\right)+a_1^2 \left((\varrho +1)^2-12
   \mathcal{R}_{22}^{h}\right)-12 (\varrho  (4 \varrho -11)+3)
   \\
   \phantom{=}&+2 \varrho  (\varrho  (2 \varrho  (\varrho
   +1)+21)+149)-80\big)+4 \mathcal{R}_{22}^{h} \left(a_1-2 \varrho +1\right) \left(252 a_1-\varrho  (25 \varrho
   +323)+80\right)\Big)
   \\
   \phantom{=}&-\left(7 a_1-\varrho  (\varrho +16)+6\right) \big(a_1 \left((\varrho +1)^2 (8
   \varrho +21)-24 \mathcal{R}_{22}^{h} \left(3 b_1 \left(b_1+2\right)-17 \varrho +10\right)\right)
e   \\
   \phantom{=}&-2 a_1^2 \left(60
  \mathcal{R}_{22}^{h}+(\varrho +1)^2\right)+3 \varrho ^2 \left(4 b_1 \left(b_1+2\right)-112 \mathcal{R}_{22}^{h}-13\right)+24
  \varrho  \left(b_1 \left(b_1+2\right) (6 \mathcal{R}_{22}^{h}+1)+17\mathcal{R}_{22}^{h}\right)
  \\
  \phantom{=}&-12 b_1 \left(b_1+2\right) (6
   \mathcal{R}_{22}^{h}-1)-120 \mathcal{R}_{22}^{h}-8 \varrho ^4-40 \varrho ^3+10 \varrho +17\big) +672 \mathcal{R}_{22}^{h} \left(a_1-2 \varrho +1\right){}^3
   \\ 
   \phantom{=}&-4 \left(7 a_1-\varrho 
   (\varrho +16)+6\right) \left(13 a_1-17 \varrho +4\right) \left(6\mathcal{R}_{22}^{h} \left(a_1-2 \varrho
   +1\right)-(\varrho +1)^2\right)\Bigg]\tilde x+\mathcal{O}(\tilde{x})\,.
    \end{split}
\end{equation}
As typically occurs in the Frobenius approach, the near-horizon series
are entirely specified in terms of a single coefficient. In this case
these are the horizon values $\mathcal{R}_{22}^{h}$ and
$\mathcal{R}_{42}^{h}$, \ie the same quantities that we need in order to
determine the power emitted in a BZ jet via Eq.~\eqref{eq:BZKRZ}. In the
Kerr case, the explicit values of $\mathcal{R}_{22}^{h}$ and
$\mathcal{R}_{42}^{h}$ can be determined analytically by means of the
Green's function method. Analogously, the asymptotic Frobenius expansions
for $\tilde{x}\to 1$, read
\begin{equation}
  \label{eq: exp2}
  \begin{split}
     \mathcal{R}_{22}&=\frac{1}{8} (\varrho
     +1)^3 \left(1-\tilde{x}\right)+
     \left[C_{22}^\infty+\frac{1}{40} (\varrho +1)^3 (10 b_1-\varrho
     -1) \log \left(1-\tilde{x}\right)\right](1-\tilde{x})^2+\mathcal{O}(1-\tilde{x})^3\, , 
     \\
     \mathcal{R}_{42}&=\frac{(\varrho +1)^5}{224
       \left(1-\tilde{x}\right)}+\frac{(\varrho +1)^2}{42}  \left[C_{22}+ (\varrho +1)^2\frac{\left(26-5 b_1 (\varrho +1)+56 \varrho  (\varrho +2)\right)}{240} + (\varrho +1)^3\frac{\left(10 b_1-\varrho -1\right)}{40}  \log
   \left(1-\tilde{x}\right)\right]
     \\
     &\frac{(\varrho +1)^2 \left(\tilde{x}-1\right)}{26880} \Big[4 b_1 \left((\varrho +1)^3 (91 \varrho +121)-720
       C^\infty_{22}\right)+980 b_1^2 (\varrho +1)^3
       \\
       &-3 (\varrho +1) \big(480 C^\infty_{22}-480 \mathcal{R}_{22}^{h}+\varrho  (\varrho
       +2) (67 \varrho  (\varrho +2)-146)+347\big)+36 (\varrho +1)^3 \left(1-10 b_1+\varrho\right) \left(2 b_1+\varrho +1\right) \log \left(1-\tilde{x}\right)\Big]
     \\
     &-(1-\tilde x)^2\Big[C_{42}^\infty+\frac{(\varrho +1)^2}{32928000} \Big(144280 b_1^3 (\varrho +1)^3+
       \\
       &+b_1 (67200
       C_{22}^\infty (5 \varrho +9)+2 (\varrho +1) (-1764000 +\varrho  (\varrho  (\varrho  (472275 \varrho
    +1840972)+1525866)-514884)+356947))
    \\
    &-8400 a_1 (\varrho +1)^2 \left(16 b_1-37 \varrho -31\right)+4 b_1^2 \left(285600 C_{22}^\infty+(271503 \varrho +62743) (\varrho
    +1)^3\right)+
    \\
    &-(\varrho +1)^2 (218400  C_{22}^\infty-352800  C_{22}^\infty\varrho 
    (\varrho  (175717 \varrho  (\varrho +4)+243702)-607532)+87517)\Big)\log(1-\tilde x)
    \\
      &-\frac{(\varrho +1)^5}{156800} \left(1-10 b_1+\varrho \right) \left(4 b_1 (5 \varrho +9)+68 b_1^2-13 (\varrho
       +1)^2\right) \log^2\left(1-\tilde{x}\right) \Big]
     +\mathcal{O}(1-\tilde{x})^3\, , 
    \end{split}
\end{equation}
\end{widetext}

Again, the series contain two unspecified parameters, respectively
labelled $C_{22}^\infty$ and $C_{42}^\infty$.

From the equation above it is immediate to see that
$\mathcal{R}_{22}^\infty=0$, whereas the function $\mathcal{R}_{42}$
presents an asymptotic divergence that cannot be removed. In fact, this
behaviour signals the presence of an outer light-surface scaling in a
non-perturbative manner, and the necessity of an analytic continuation of
the function beyond the outer light surface via a matched asymptotic
expansion~\cite{Armas2020, Camilloni2022}.

Since our aim is to explicitly compute the BZ power at the horizon, 
Eq.~\eqref{eq:BZKRZ}, we are only interested in the values of the
functions at the horizon, and such a divergence for $\mathcal{R}_{42}$
does not play any role in the present discussion. Without loss of
generality we can thus make the computation easier by defining a new
function
\begin{equation}
  \label{eq:barR}
  \bar{\mathcal{R}}_{42}(\tilde x) := \mathcal{R}_{42}(\tilde
  x)+\frac{1}{224}\frac{(\varrho +1)^5}{1-\tilde{x}}\, , 
\end{equation}
that is finite in the entire integration domain. Notice that its horizon
value is simply rescaled via
\begin{equation}
\bar{\mathcal{R}}^{h}_{42} = \mathcal{R}_{42}^{h}+\frac{(\varrho
  +1)^5}{224}\,.
\end{equation}
At this point, we can use the shooting method to solve the ODEs and
determine $\mathcal{R}_{22}^{h}$ and $\mathcal{R}_{42}^{h}$.
In particular we construct a ``left solution'',
$\mathcal{R}^{\rm L}(\tilde{x})$, by integrating the equations from the
horizon $\tilde x=0$ up to a matching point $\tilde{x}_m$, where the
horizon boundary condition is consistent with Eq.\eqref{eq: exp1}. The
left solutions, thus, depend on the near-horizon shooting coefficients
$\mathcal{R}^{h}_{22}$ and $\mathcal{R}^{h}_{42}$.
Analogously, a ``right solution'', $\mathcal{R}^{\rm R}(\tilde{x})$, can be
constructed by integrating the equations from the infinity $x=1$ down to
a matching point $\tilde{x}_m$, where we set an asymptotic boundary
conditions consistent with the expansion~\eqref{eq: exp2}. The right
solutions depend on the asymptotic shooting coefficients $C^\infty_{22}$
and $C^\infty_{42}$.
\begin{align}
    &\mathcal{R}^{h}_{22}(0) \!-\! \left(\frac{6\pi^2 \!-\!
    49}{72}\right)\sim 10^{ \!-\! 17}\, , \\ \nonumber
  \\ &\mathcal{R}^{h}_{42}(0) \!-\!  \left(\frac{39\zeta(3)}{3920} \!+\!
  \frac{17929399}{2540160} \!-\!  \frac{3877\pi^2}{12096} \!-\!
  \frac{19\pi^4}{480}\right) \!\sim\! 10^{ \!-\! 16}\, .
\end{align}
We expect that a unique solution can be pinned out by requiring the
consistency between the left and right solutions, \ie the consistency of
the near-horizon and asymptotic expansion close to the boundaries.
The shooting parameters regulating the boundary values, 
$\mathcal{R}^{h}$ and $C^\infty$, are therefore varied until the
matching condition
\begin{equation}
  \begin{split}
    \mathcal{R}^{\rm L}(\small{\mathcal{R}^{h}};\tilde{x}_m) & =
    \mathcal{R}^{\rm R}(C^\infty;\tilde{x}_m)\, ,
    \\ \partial_{\tilde{x}}\mathcal{R}^{\rm
      L}(\small{\mathcal{R}^{h}};\tilde{x}_m) &
    =\partial_{\tilde{x}}\mathcal{R}^{\rm R}(C^\infty;\tilde{x}_m)\, ,
  \end{split}
\end{equation}
is satisfied for a certain combination of $\mathcal{R}^{h}$ and
$C^\infty$. The integration is also repeated by scanning across the
parameter space $(\varrho, a_1, b_1)$. We verified that the result does
not depend on the choice of the matching point $\tilde{x}^m$.
The final results of the integration via shooting are represented in
Fig.~\ref{fig1}. The values of $\mathcal{R}_{22}^{h}$ and
$\mathcal{R}_{42}^{h}$ as a function of $\varrho$, $a_1$ and $b_1$ are
shown together with insets illustrating some examples of the matching
between the horizon and the asymptotic solutions.
As a sanity check for our method we verified that for
$(\varrho, a_1, b_1)=0$ the known analytic values for a monopolar Kerr
magnetosphere (see Eq.(4.22a) in Ref.~\cite{Camilloni2022}) are
reproduced with great accuracy
The Kerr values above can be used to give approximate analytic
expressions for $\varrho, a_1, b_1 \ll 1$. We found that in the range
$\varrho, a_1, b_1\in [-0.1, 0.1]$ the function are well approximated by
with the following quadratic polynomials, obtained via fitting
\begin{align}
  \label{eq:R22R42appr}
  \mathcal{R}^{h}_{22}(\varrho)&\approx\mathcal{R}^{h}_{22}(0)+0.647
  \varrho +1.458 \varrho^2\, ,
  \\ \mathcal{R}^{h}_{22}(a_1)&\approx\mathcal{R}^{h}_{22}(0)-0.1348 a_1
  +0.1272 a_1^2\, ,
  \\ \mathcal{R}^{h}_{22}(b_1)&\approx\mathcal{R}^{h}_{22}(0)-0.0319b_1
  -0.0268 b_1^2\, ,
  \end{align}
and
\begin{align}
  \mathcal{R}^{h}_{42}(\varrho)&\approx\mathcal{R}^{h}_{42}(0)+0.0794
  \varrho -0.547 \varrho^2\, ,
  \\ \mathcal{R}^{h}_{42}(a_1)&\approx\mathcal{R}^{h}_{42}(0)+0.01442 a_1
  -0.099 a_1^2\, ,
  \\ \mathcal{R}^{h}_{42}(b_1)&\approx\mathcal{R}^{h}_{42}(0)-0.0139b_1
  +0.0055 b_1^2\, ,
  \label{eq:R22R42appr_last}
\end{align}
The values reported in Fig.~\ref{fig1}, as well as the approximations in
Eqs.~\eqref{eq:R22R42appr}--\eqref{eq:R22R42appr_last}, have been used in
the main text to compute the BZ luminosity in Eq.~\eqref{eq:BZKRZ}.

\newpage

\bibliography{aeireferences}

\end{document}